\documentclass[aps,prb,superscriptaddress,twocolumn]{revtex4-1}
\usepackage{amsmath}
\usepackage{ amssymb }
\usepackage{graphicx}
\usepackage{color}

\begin{document}

\newcommand{\beq}{\begin{equation}}
\newcommand{\eeq}{\end{equation}}
\newcommand{\beqa}{\begin{eqnarray}}
\newcommand{\eeqa}{\end{eqnarray}}
\newcommand{\ben}{\begin{enumerate}}
\newcommand{\een}{\end{enumerate}}
\newcommand{\hs}{\hspace{0.5cm}}
\newcommand{\vs}{\vspace{0.5cm}}
\newcommand{\note}[1]{{\color{red} \bf [#1]}}
\newcommand{\tim}{$\times$}
\newcommand{\bigo}{\mathcal{O}}

\title{Corner contribution to the entanglement entropy \\ of an $O(3)$ quantum critical point in 2+1 dimensions}

\author{Ann B.\ Kallin}
\affiliation{Department of Physics and Astronomy, University of Waterloo, Ontario, N2L 3G1, Canada}

\author{E.M.\ Stoudenmire}
\affiliation{Perimeter Institute for Theoretical Physics, Waterloo, Ontario, N2L 2Y5, Canada}

\author{Paul Fendley}
\affiliation{Physics Department, University of Virginia, Charlottesville, VA 22904-4714 USA}

\author{Rajiv R.P.\ Singh}
\affiliation{Physics Department, University of California, Davis, CA, 95616 USA}

\author{Roger G.\ Melko}
\affiliation{Department of Physics and Astronomy, University of Waterloo, Ontario, N2L 3G1, Canada}
\affiliation{Perimeter Institute for Theoretical Physics, Waterloo, Ontario, N2L 2Y5, Canada}

\date{\today}

\begin{abstract}
The entanglement entropy for a quantum critical system across a boundary with a corner exhibits a sub\-leading logarithmic scaling term with a scale-invariant coefficient. 
Using a Numerical Linked Cluster Expansion, we calculate this universal quantity for a square-lattice bilayer Heisenberg model
at its quantum critical point. 
We find, for this $2+1$ dimensional $O(3)$ universality class, that it
is thrice the value calculated previously for the Ising universality class. This relation gives substantial evidence that this coefficient provides a measure of the number of degrees of freedom of the theory, analogous to the central charge in a $1+1$ dimensional conformal field theory.

\end{abstract}

\maketitle

\section{Introduction}

Entanglement between two subregions of a system provides a novel probe of quantum correlations.\cite{EEreview}
For this probe to be useful, it is necessary to extract quantities that are not only universal, but also give intuition into physical properties. 
One prominent idea is to use the entanglement entropy to define a measure of the degrees of freedom at and near quantum critical points. Heuristically, such a measure quantifies the information lost during RG transformations so that one may constrain flows of theories relevant for many condensed matter systems.\cite{Casini12,Grover_C}
While much is known for $1+1$ dimensional systems via the connection of Zamolodchikov's $c$-theorem \cite{Zamo} to 
entanglement,\cite{Holzhey,VLRK,Korepin,Cardy} studies in higher dimensionality are in their infancy.

Subleading terms in the entanglement entropy provide some very intriguing possibilities. The leading contribution to the entanglement entropy $S(A)$, between two subregions $A$ and $B$, scales
with the ubiquitous area law \cite{Sorkin,Shredder,ALreview}
(with a few important exceptions\cite{Wolf2,Israel,EBL}).
In a quantum critical system in $d+1$ space-time dimensions, a shape-dependent subleading term $\gamma$ is expected, so that \cite{Fursaev}
\begin{equation}
S(A) = C \left({ \frac{\ell}{\delta} }\right)^{d-1}\ +\ \dots\ +\ \gamma\ +\ \cdots. \label{AreaLaw}
\end{equation}
The linear dimension $\ell$ characterizes the subregion $A$ and $\delta$ is the ultraviolet/lattice-scale cutoff, while
the ellipses conceal non-universal constants and subleading terms depending on the length scale to some power.
Explicit expressions for $\gamma$ in a cylindrical geometry with smooth boundaries were found for $\phi^4$ theory in $d=3-\epsilon$.\cite{Max} At the infrared-unstable free-field fixed point, 
$\gamma$ indeed decreases under the renormalization group flow toward the non-trivial Wilson-Fisher fixed point, just like the $c$-theorem requires for the central charge $c$ in $d=1$. \footnote{For a simple explanation of how $c$ measures degrees of freedom in 1+1 dimensions, see Ref. 59.}

The precise functional form of $\gamma$ is expected in general to depend on scale invariants 
such as the aspect ratios, Euler characteristic, and other geometric features of region $A$.\cite{Fradkinbook} 
A particularly interesting piece comes from terms involving $\log(\ell/\delta)$. With smooth boundaries, this logarithmic divergence occurs only for odd $d$.\cite{ryu,ryu_2} For even $d$, it can occur when there is a singularity such as a corner or a cone in the surface separating $A$ and $B$. 
In $d=2$, the case of interest here,  the contribution to $\gamma$ of a corner of interior angle $\theta$ in the one-dimensional boundary is
\begin{equation}
\gamma = a(\theta)\log\left({  \frac{\ell}{\delta}   }\right) + \cdots .
\end{equation}
Since the cutoff length $\delta$ is contained within the logarithm, rescaling it only affects the terms in the ellipses, not the coefficient $a(\theta)$. 

Such corner contributions in $d=2$ have been computed in several interesting situations. At a conformal quantum critical point  \cite{AFF} such as the quantum Lifshitz theory (describing e.g.\ the square-lattice quantum dimer model \cite{RK}), results from two-dimensional conformal field theory \cite{cardy-peschel} have been adapted to give $a(\theta)$.\cite{FradkinMoore} It has been calculated in free-scalar field theory,\cite{logcorner} numerically in interacting lattice models,\cite{Kallin_Heis,TFIM_series,Kallin_NLCE,Inglis_QMC}
and more generally using the AdS/CFT correspondence.\cite{Hirata} These are consistent with a conjectured geometrical form valid in any dimension.\cite{Robcorner}

These computations confirm that $a(\theta)$ is indeed universal. Even more strikingly, they
point to $a(\theta)$ as a useful measure of the number of degrees of freedom. For example, for a $2+1$-dimensional conformal quantum critical point, $a(\theta)$ is proportional to the central charge $c$ of the two-dimensional conformal field theory describing the ground-state wavefunction.\cite{FradkinMoore}

The purpose of this paper is to show that this corner contribution behaves similarly in a less exotic situation, namely an 
interacting theory familiar to both condensed matter and particle theorists. 
To this end, we study the entanglement entropy for a critical system of spin-$1/2$ particles on a square-lattice bilayer with nearest-neighbor antiferromagnetic Heisenberg interactions. This quantum critical point is in the same universality class as the three-dimensional classical Heisenberg model,\cite{Wang_bilayer} described in the continuum limit by the well-studied three-component $O(3)$-invariant $\phi^4$ field theory.\cite{Janke}

Specifically, we numerically compute the universal coefficient $a(\theta=\pi/2)$ in a continuous family of Renyi entanglement entropies for this critical Heisenberg bilayer. We compare the results to those for the critical point of the transverse-field Ising model (TFIM),\cite{Kallin_NLCE} whose field theory description is a single-component $\phi^4$ theory. We find over a wide range of Renyi index values, $a(\pi/2)$ for the former is thrice that for the latter, to within numerical uncertainties. The factor of three is compelling evidence that the universal coefficient of the corner-induced logarithm provides a measure of the number of low-lying degrees of freedom.
This indeed is behavior analogous to that of $c$ in $1+1$ dimensional CFTs.  

In an interacting theory in general $d$, one does not expect such a measure to literally count the degrees of freedom; rather, it provides a way of understanding which renormalization-group flows are possible. The factor of 3 we find is presumably a consequence of the fact that $O(N)$ $\phi^4$ theory is ``close'' to free, in that the $\epsilon$ expansion around a free-field theory applies. To be precise, our numerical result implies that the leading contribution to $a(\theta)$ in $N$-component $\phi^4$ field theory in $2+1$ dimensions is proportional to $N$, and that corrections are within our (fairly small) numerical error.

The outline of this paper is as follows.
In section \ref{sec:bilayer} we review the Heisenberg bilayer model and its theoretical description. In section  \ref{sec:NLCE}, we describe our numerical method, a novel Numerical Linked-Cluster Expansion (NLCE).\cite{NLC1,NLC2,NLC3,NLC4}
Our NLCE uses both Lanczos diagonalization and Density Matrix Renormalization Group\cite{White92,Scholl05,Stoudenmire} (DMRG) simulations to calculate the Renyi entanglement entropies.\cite{A_renyi} We present our results in section \ref{sec:results}, and discuss some implications of this work in section \ref{sec:conclusion}.

\section{The Heisenberg Bilayer Model and Its Critical Point}
\label{sec:bilayer}

\begin{figure}[t]
\includegraphics[width=3.6in]{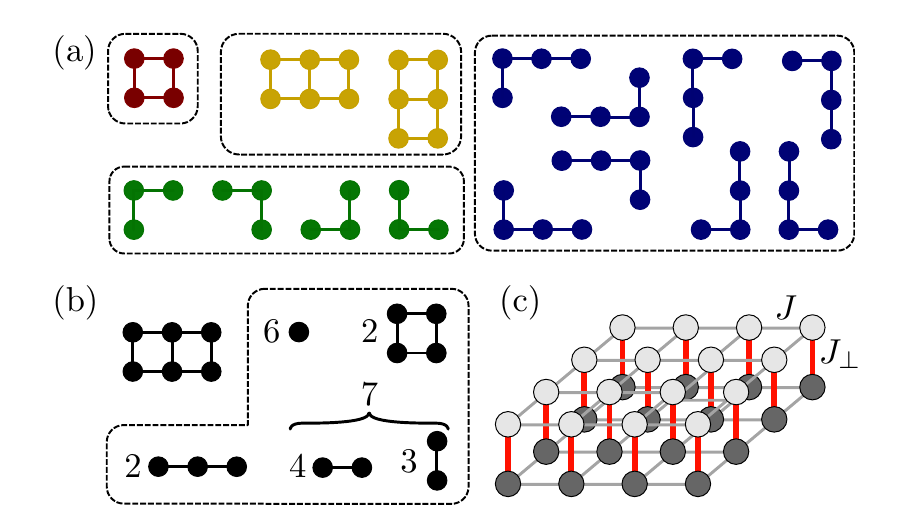}
\caption[]{
(a) The lattice constant $L(c)$ denotes the number of distinct ways a cluster can be embedded in a lattice.  
The boxes illustrate the four possible values of $L(c)=1,2,4,8$ for a square lattice.
(b) The subclusters of a $2\times3$ cluster, corresponding to row (6) of Table \ref{table1}.
(c) A $4\times4\times2$ bilayer cluster (row (10) in Table \ref{table1}) with intralayer coupling $J$ denoted by the thin grey bonds, and interlayer coupling $J_{\perp}$ denoted by the red vertical bonds.
\label{latt_const}
}\end{figure}

The aim of this paper is to understand the universal subleading term coming from corner contributions to the entanglement entropy. We study a strongly interacting quantum model whose physics is well understood, and which is amenable to treatment by the powerful NLCE method described in section \ref{sec:NLCE}. This system is the spin-1/2 nearest-neighbor Heisenberg model on a square lattice bilayer, or equivalently, two flavors of spins on the square lattice. Labelling the spin operator on site $j$ and layer $a$ by ${\bf S}_{aj}$,
the Hamiltonian is
\begin{equation}
H = J \sum_{\langle i,j \rangle} \left({ {\bf S}_{1i} \cdot {\bf S}_{1j}  + {\bf S}_{2i} \cdot {\bf S}_{2j} }\right) + J_{\perp} \sum_i {\bf S}_{1i} \cdot {\bf S}_{2i}, \label{eq:ham}
\end{equation}
where $\langle i,j \rangle$ denotes a pair of nearest-neighbor sites within a layer. We take the couplings $J$ and $J_\perp$ as positive, hence antiferromagnetic.
The interactions between sites on the same layer ($J$) and sites on different layers ($J_\perp$) are represented pictorially in Fig.~\ref{latt_const}(c) for a 4\tim4\tim2 bilayer system.

This Hamiltonian supports two different phases in its ground-state. At $J_\perp=0$, the physics is simply that of two decoupled square-lattice Heisenberg models. These are N\'eel ordered, so that the $SU(2)$ symmetry is spontaneously broken to $U(1)$.  For small $J_{\perp}/J$, the order persists, and the coupling between the layers relates the ordering. Thus there are two Goldstone bosons in this N\'eel phase
(which, incidentally, give rise to a subleading logarithm in the entanglement entropy of a straight boundary on a finite-size lattice\cite{Kallin_Heis,Max_Tarun}). 
At large $J_{\perp}/J$, there is a ``dimer'' phase where the pair of spins on each bond between the bilayers forms a singlet. In this phase the $SU(2)$ symmetry is unbroken, and the spectrum is gapped. 
A single quantum critical point separates the two phases.
Estimates for the critical coupling have been calculated to high accuracy with 
series expansion \cite{Zheng_bilayer,Hamer_bilayer} and quantum Monte Carlo, \cite{Wang_bilayer} with the most accurate estimate $( J_{\perp} / J)_c = 2.5220(2)$ coming from the latter.

The universality class of this phase transition turns out to be the simplest possibility consistent with the symmetry. The order parameter describing the N\'eel/Goldstone phase is the staggered magnetization.
Following the Landau-Ginzburg approach, one defines a three-component vector field $\vec{\phi}$ representing a suitably averaged order parameter. The simplest action for this field consistent with the symmetries is the $O(3)$-invariant $\phi^4$ theory:
\begin{equation}
S=\int d^2x dt \left( \frac{\partial\vec\phi}{\partial t}\cdot \frac{\partial\vec\phi}{\partial t}- \nabla\vec\phi\cdot\nabla\vec\phi - \mu^2\vec\phi\cdot\vec\phi - g (\vec\phi\cdot\vec\phi)^2 \right).
\label{eq:action}
\end{equation}
At $\mu^2<0$ the $O(3)$ symmetry is spontaneously broken to $SO(2)$, resulting in two Goldstone bosons.  There is a continuous phase transition at $\mu^2=0$ to a phase with no order in $\vec\phi$, as in the bilayer. 

It has long been known that this Landau-Ginzburg/effective field theory describes the phase transition in the three-dimensional classical Heisenberg magnet between the low-temperature Goldstone and the high-temperature disordered phases. Even though the underlying degrees are fixed-length spins (and so are labeled by two angles), the average value of the magnetization also fluctuates strongly near the critical point, and so this three-component theory describes the critical behavior. This has been confirmed by comparing detailed numerical simulations \cite{Janke,Campo} with calculations in the $d=3-\epsilon$ expansion.\cite{ZJ}

The same Landau-Ginzburg approach applies to the Heisenberg bilayer model we study here, with the same conclusion. This has also been confirmed convincingly via numerics,\cite{Wang_bilayer} so that 
critical fluctuations corresponding to the restoration of $SU(2)$ symmetry at the critical point indeed are described by three bosonic fields. The appearance of the longitudinal mode can also be understood directly in the quantum theory.\cite{Chubukov_bilayer} It is a bound state in the Ising limit and therefore neglected by spin-wave and Schwinger boson treatments, 
which among other things leads to very poor estimates for the critical coupling $( J_{\perp} / J)_c$. This situation can be remedied with a proper treatment of
all three modes\cite{bilayer_SWT1,bilayer_SWT2,Sommer}).
The existence of this additional longitudinal mode has been recently emphasized by a numerical study explicitly demonstrating that it becomes degenerate with the two Goldstone modes exactly at the critical point, corresponding to a full restoration of $SU(2)$ symmetry.\cite{Hamer_bilayer}

Thus, the quantum critical point of the Heisenberg bilayer model offers an excellent opportunity to examine the behavior of the 
entanglement entropy in the presence of multiple bosonic modes in the low-lying spectrum of an interacting model.In the next section, we discuss the details of the numerical simulation scheme used to extract the corner contribution to the
Renyi entanglement entropies.

\section{Numerical Linked-Cluster Expansion}
\label{sec:NLCE}

The Numerical Linked-Cluster Expansion\cite{NLC1,NLC2,NLC3,NLC4} (NLCE) is a method of extending measurements of a series of finite-sized lattice clusters towards the thermodynamic limit, cancelling off finite-size and boundary effects using sums and differences of various clusters.
In its general formulation, this method uses measurements of a given property $P$ from all possible clusters of sites that can be embedded in the chosen lattice.  
Typical NLCE approaches involve the computationally expensive task of generating all clusters, and embeddable subclusters.
This cluster embedding problem results in an exponential bottleneck, restricting
the maximum number of sites to $\sim 16$ or so, depending on the lattice.
However, in Ref.~\onlinecite{Kallin_NLCE}, a modified NLCE procedure was defined that employs an alternative definition
of cluster geometries, involving only $m \times n$ rectangles.  This restriction significantly simplifies the cluster embedding
problem, passing the computational bottleneck to the calculation of ground-state properties, via numerical techniques such
as exact diagonalization (Lanczos) or DMRG.  As a consequence, one is able to achieve significantly higher orders of the expansion than 
conventional NLCE, and therefore significantly improve approximations to the thermodynamic limit. In this section, we give details of the 
NLCE procedure, definition of cluster geometries for the Heisenberg bilayer model, methods to define order extrapolations, and
procedures to calculate the Renyi entanglement entropies using the Lanczos and DMRG methods.

\subsection{NLCE Overview} \label{overview}
The foundation of the NLCE method is based on the fact 
that properties of a lattice model can be expressed as a sum over all distinct clusters which are embeddable in the lattice ${\cal L}$. Let a cluster $c$ be a set of sites with the connectivity of the underlying lattice, and 
$L(c)$ be the number of distinct ways it can be embedded. A rotation or reflection could result in a different embedding depending on the symmetry of $c$, whereas a simple translation will not lead to a different embedding in the infinite lattice $\mathcal{L}$; see Figure \ref{latt_const}(a)).  For a square lattice, $L(c)$ can take values 1, 2, 4 and 8. Here, we follow Ref.~\onlinecite{Kallin_NLCE} and consider only $m \times n$ rectangular clusters, for which 
$L(c)$ can only be equal to 1 (if $m = n$) or 2 (if $m \neq n$).

A property $P$ per site can then be expressed as
\beq
P({\cal L})/N = \sum_c L(c) \times W(c)\ ,
\label{NLCeq1}
\eeq
where the weight of a cluster for a given $P$ is defined by
\beq
W(c) = P(c) - \sum_{s\in c} W(s),
\label{NLCeq2}
\eeq
where $s$ is any subcluster of $c$. 
In the sum, each subcluster $s$ is included the number of times that it can fit inside cluster $c$ (see Figure \ref{latt_const}(b)).
This relation is a generalization of the {inclusion-exclusion} principle, the statement that the number of elements in the union of two finite sets $|A \cup B|$ is simply $|A|+|B|-|A\cap B|$. This gives intuition for why disconnected clusters need not be considered in NLCE calculations:
their weight $W(c)$ is always zero, just as the number of elements in the intersection of disconnected sets is zero.

The NLCE procedure uses Eqs.~\ref{NLCeq1} and \ref{NLCeq2} to build up 
the value of $P/N$, starting from the smallest cluster (which has no subclusters, thus $W(1) = P(1)$) 
and ending with some maximal cluster size.  For rectangular clusters on a square lattice, this maximal cluster size is
limited in practice only by the computational cost of calculating the given property $P$ on the cluster.\cite{Kallin_NLCE}

The value of the weight $W(c)$ is an indicator of the convergence of the NLCE.
In a system with no broken symmetries and a finite correlation length, the weights should decrease exponentially
with cluster size, once these sizes exceed the correlation length. This would lead to an exponential convergence
of the NLCE with cluster size (or ``order'' as we define below). At (and near) a critical point, where the correlation length 
becomes large compared to the sizes of clusters we can study, 
the weights will vary as a power of the cluster size. This will lead to an algebraic convergence with order for
quantities like ground-state energy and entanglement entropies, requiring a careful extrapolation. Note that
other quantities, such as order-parameter susceptibility 
will diverge at the critical point. But even for divergent properties, the NLCE can extract useful information if 
one can reach orders large enough for the property to be fit to a known scaling relation.
The fact that some quantities converge  and others diverge
is analogous to convergence or divergence of a series expansion at its radius of convergence.

It is instructive to work out the NLCE for one-dimensional systems, which can be done
in full generality.
All sub-clusters of a one-dimensional system are uniquely labeled by their length $n$, and
can be embedded only one way, thus $L(n) \equiv 1$ and
\begin{align}
P/N = \sum_{n=1}^{\infty} W(n) \ .
\end{align}
The cluster weights are
\begin{align}
W(1) & = P(1) \\
W(2) & = P(2)-2 P(1) \\
W(3) & = P(3)-2 P(2) + P(1) \\
\cdots \nonumber \\
W(n) & = P(n)-2 P(n\!-\!1) + P(n\!-\!2) \ \ \ [n \geq 3].
\end{align}
Defining partial sums up to order $n$ of the NLCE as
\begin{align}
p(n) = \sum_{m=1}^{n} W(m) \ ,
\end{align}
then for $n > 1$
\begin{align}
p(n) = P(n) - P(n-1) \:. \label{eqn:partial_sum}
\end{align}
Assuming that $P(n)$ changes ever more slowly with increasing $n$, then
in the limit of large clusters
\begin{align}
W(n) & \simeq \frac{\partial^2}{\partial n^2} P(n) \\
p(n)  & \simeq \frac{\partial}{\partial n} P(n) \:.
\end{align}
It is interesting to observe from Eq.~(\ref{eqn:partial_sum}) that in $1d$, the NLCE is
nothing but the ``subtraction trick'' used, for example, in DMRG calculations to estimate bulk
properties in the thermodynamic limit from finite systems with open
boundary conditions.\cite{Stoudenmire}

\subsection{Rectangular Clusters for the Square Lattice Bilayer}
\label{clusters}

\begingroup
\begin{table}
	\begin{tabular}{| c | c | c | c | c | c | c | c | c | c | c | c | c | c |}
    		\hline
		\multicolumn{4}{|c|}{} && \multicolumn{9}{|c|}{Subcluster Multiplicity} \\ \hline 
		
		id\# & $n_x \times n_y$ & $N$ & $L(c)$&&(1) & (2) & (3) & (4) & (5) & (6) & (7) & (8) & (9) \\ \hline \hline
		(1) & $1 \times 1$ & 2 & 1 &&&&&&&&&&\\ \hline
		(2) & $1 \times 2$ & 4 & 2 &&2&&&&&&&&\\ \hline
		(3) & $1 \times 3$ & 6 & 2 &&3&2&&&&&&&\\ \hline
		(4) & $1 \times 4$ & 8 & 2 &&4&3&2&&&&&& \\ \hline
		(5) & $2 \times 2$ & 8 & 1 &&4&4&&&&&&&\\ \hline
		(6) & $2 \times 3$ & 12 & 2 &&6&7&2&&2&&&& \\ \hline
		(7) & $2 \times 4$ & 16 & 2 &&8&10&4&2&3&2&&&\\ \hline
		(8) & $3 \times 3$ & 18 & 1 &&9&12&6&&4&4&&&\\ \hline
		(9) & $3 \times 4$ & 24 & 2 &&12&17&10&3&6&7&2&2&\\ \hline
		(10) & $4 \times 4$ & 32 & 1 &&16&24&16&8&9&12&6&4&4\\ \hline
	\end{tabular}
	\caption[]{
The properties of all clusters up to a maximum $n_x$ and $n_y$ of four sites, including: number of sites $N$, lattice constant $L(c)$, and the multiplicity of their subclusters (this is left blank when the multiplicity is zero).
There is an implied bilayer ($n_z=2$) for each cluster.
\label{table1}
}
\end{table}
\endgroup

In this paper we study a square lattice bilayer system.  Each cluster used in the calculation is an $n_x\times n_y \times 2$ array of sites, where $n_x,n_y \ge 1$.
The second layer of the bilayer lattice does not change the NLCE calculations from the strictly two dimensional ($2d$) NLCE, other than doubling the number of sites in a given cluster, increasing the computational expense of measuring most properties. Table~\ref{table1} shows the characteristics of all rectangular clusters with $n_x, n_y \le 4$, including the number of embeddings of each subcluster, needed for Eq.~\eqref{NLCeq2}.
Unfortunately the $2d$ cluster weights do not simplify as in the $1d$ case, and all terms (even those from the smaller clusters) 
need to be included in Eq.~\eqref{NLCeq1}. 

\begin{figure}[t]
\includegraphics*[width=2.3in]{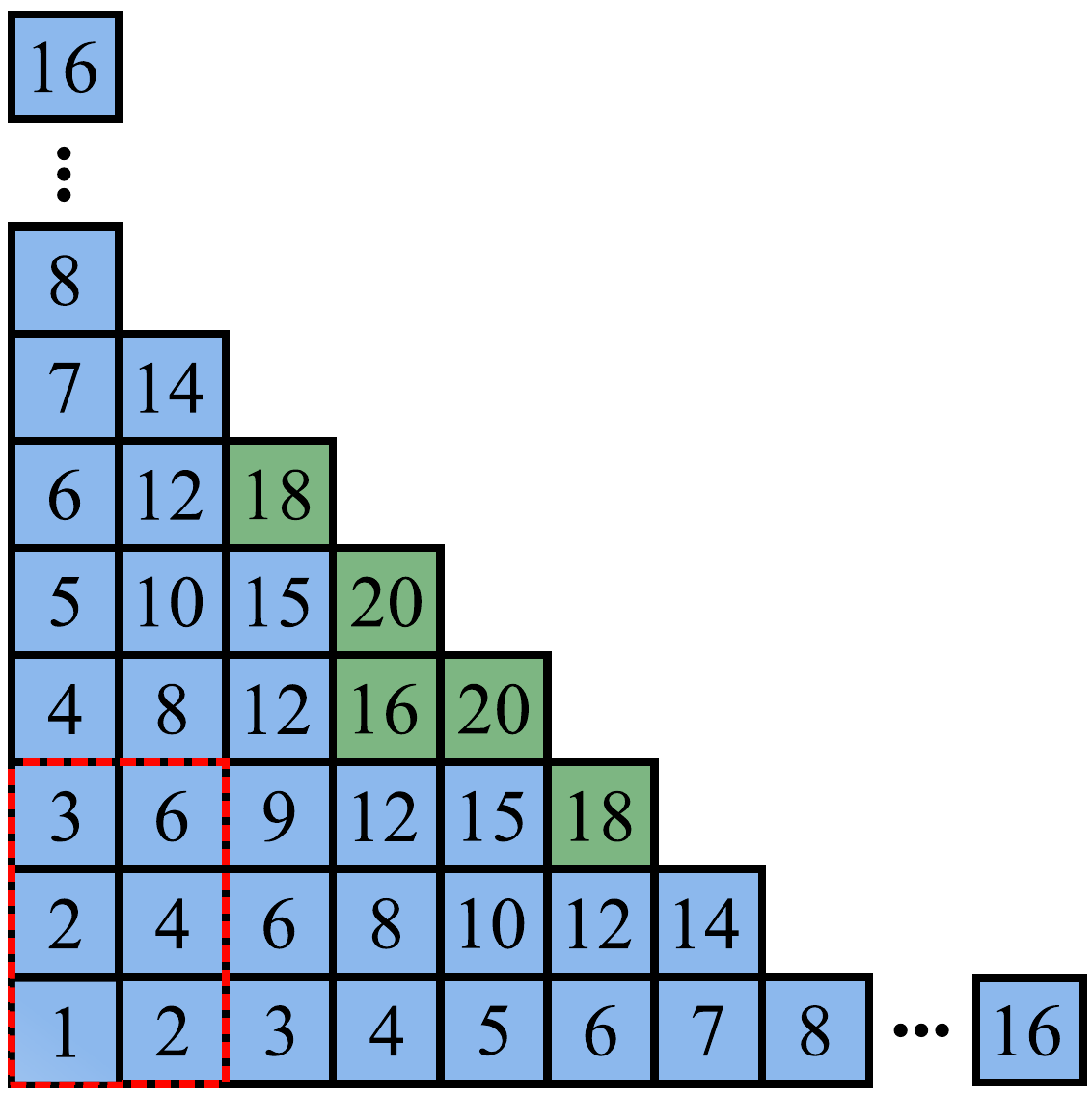}
\caption[]{
Two methods for defining order $\bigo$.  Each tile shows the number of sites contained in a rectangle with its top right corner in that square and its bottom left corner at the bottom left of the figure.  As an example the 2\tim3 rectangle is outlined with a dashed line, and from the top right corner one can see that it contains 6 sites per layer.  Using $\bigo_1$ the different orders are defined by the diagonals of this diagram.  $\bigo_1=1$ contains only the 1\tim1\tim2 cluster (denoted by ``1'' in the diagram), $\bigo_1=1.5$ contains the clusters denoted by the 2's, $\bigo_1=2$ contains 3,4,3 and so forth, moving along the diagonals.
$\bigo_2$ is simple to understand from this diagram.  If $\bigo_2=x$, that order contains clusters with $x$ sites per layer. 
The shaded tiles show clusters solved by Lanczos and DMRG in this work.
\label{orders}
}
\end{figure}

In an NLCE calculation, it is necessary to define an {\em order}, a way of grouping together clusters of similar sizes.
This allows one to assign a length scale related to the largest order included in a calculation.
At a quantum critical point for example, this allows one to study NLCE data as a function of order,
giving scaling relationships that can be extrapolated towards the thermodynamic limit.

In this paper we consider two methods of grouping clusters: (1) by the average edge-length of a cluster $\bigo_1 = (n_x + n_y)/2$ and (2)  by the number of sites $\bigo_2 = N/2$ in one layer of the cluster. 
In the two cases we have the average cluster length, $\ell \sim \bigo_1 = (n_x + n_y)/2$ and $\ell \sim \sqrt{\bigo_2} = \sqrt{n_xn_y}$, 
which can be thought of respectively as the arithmetic and geometric mean of the edge-lengths for clusters of that order; Fig.~\ref{orders} attempts to give some intuition for the two different definitions.
Using $\bigo_2$ tends to include longer $1d$ clusters, while only including smaller square clusters, whereas truncating using $\bigo_1$ includes all the clusters that will fit inside a diamond of a given size defined by that order, thus excluding long $1d$ clusters.
Additionally, each order using $\bigo_1$ includes an increasing number of clusters (equal to $\bigo_1$ rounded down to the nearest integer), while the number of clusters for each new order of $\bigo_2$ is generally smaller and determined by the number of factors of $\bigo_2$.  This tends to give the data a step-like distribution as a function of $\bigo_2$ while allowing for higher orders to be calculated, resulting in more data points.  Contrastingly, plotting data as a function of $\bigo_1$ gives a smoother distribution with fewer overall data points.  Both methods contain the same information (though $\bigo_1$ excludes long $1d$ clusters) simply viewed in different ways.

The lattice constant, as described in Section \ref{overview}, reflects the number of ways a cluster can be embedded in the underlying lattice, and in the case of our square bilayer lattice, is limited to the values $L(c) = 1,2$.
It is important that measurements considered on a cluster obey the same symmetry, i.e.~if $L(c) =2$ then a measurement must give the same results for both orientations of the cluster, otherwise the two orientations should be considered separately with $L(c)=1$ for each.
Standard measurements, such as energy, would never depend on the orientation of the cluster.
However we will see below that this becomes important for simplifying the measurement of entanglement, which depends on spatial boundaries defined in the lattice.

\subsection{The Renyi Entanglement Entropies} \label{RenyiSection}

\begin{figure}
\includegraphics*[width=3.3in]{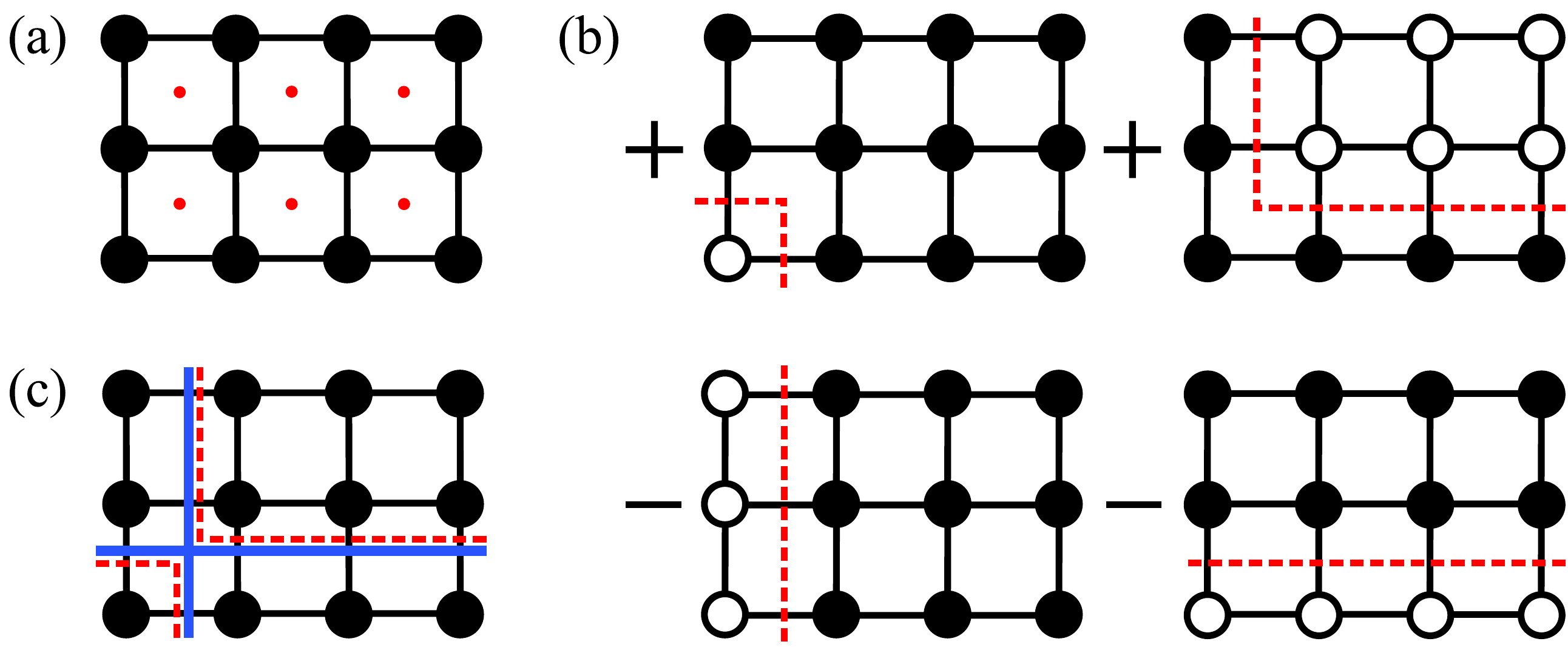}
\caption[]{
(a) The entanglement due to a corner is calculated by considering each plaquette of a cluster.  In this figure we consider a 4\tim3 cluster.
(b) The corner entanglement is equal to the difference of the entanglement due to opposing boundaries with corners ($V^1$, $V^2$) at the given plaquette and  the entanglement across to the corresponding horizontal and vertical lines ($L^x$, $L^y$).
(c) This figure shows all four entanglement cut geometries for the first plaquette superimposed.  The subtracted line cuts are solid, and the corner cuts are dotted.
\label{entanglement}
}
\end{figure}

The Renyi entanglement entropies\cite{A_renyi} are used to measure bipartite entanglement between two spatial regions of a system,
labelled $A$ and $B$,
\beq
S_{\alpha}(A) = \frac{1}{1-\alpha} \log {\rm{Tr}}(\rho_A^{\alpha}),
\eeq
where $\rho_A = {\rm Tr}_B (\rho)$ is the reduced density matrix of region $A$.  Here, $\alpha$ is the Renyi index, 
such that the limit $\alpha = 1$ gives the familiar von Neumann entropy.

As mentioned above, the measurement of Renyi entropies in NLCE is non-standard because it requires the definition of 
two spatial regions $A$ and $B$.  
In this paper we focus on the entanglement due to a $90^\circ$ corner in the boundary between entangled regions. 
This is extracted by defining a boundary with a corner, and subtracting off the entanglement contribution
coming from the linear portions, leaving only the entanglement due to the corner (Fig.~\ref{entanglement}).

Entanglement measurements in general are done by considering every possible way that the chosen boundary can intersect a cluster. 
Thus, to measure the entanglement across a line running in the $y$ direction $L^y$, for a given cluster we use
\beq
P(c) = \sum_{i=1}^{n_x-1} S_\alpha(L^y_i),
\eeq
where $L^y_i$ denotes a region $A$ including $i$ columns of the cluster.  
This measurement can be simplified for many rectangular clusters.  Since $S(A) = S(B)$ for a pure ground-state wavefunction, we have $S(A_{i}) = S(A_{n_x-i})$ which can approximately halve the number of measurements required.
Different types of boundaries will have different symmetries that can be exploited to reduce the total number of measurements required.

As mentioned in Section \ref{clustersym}, if a cluster has lattice constant $L(c) > 1$, any measurements on that cluster must share the same symmetry, or else $L(c)$ must be modified for that cluster.   
This becomes especially important  to consider for entanglement measurements, where a cluster is divided into two spatial regions.
The example above, of the entanglement across a vertical line will give different results for the two orientations of a non-square cluster.  This can be remedied by instead calculating the sum of the entanglement due to a vertical line and that due to a horizontal line.
A second, but equivalent, method would be to always treat $n \times m$ clusters as distinct from $m \times n$ clusters and measure entanglement across only a line in the $x$ direction, for example.  Both techniques require the same computational effort, but use different methods of bookkeeping for the clusters and measurements.

The NLCE, unlike many other numerical techniques, is able to isolate the entanglement due to a $90^\circ$ corner by subtracting off the entanglement contributions from the linear portions of the boundary.
This is done by subtracting the entanglement due to the lines $L^x$ and $L^y$, from entanglement across opposing lines with $90^\circ$ vertices ($V^1$ and $V^2$), as shown in Figure \ref{entanglement}(b) and (c).
To fully cancel off any line contributions to the entanglement we calculate the corner entanglement $\mathcal{V}_\alpha$ of a cluster $c$ using 
\beqa
\label{eqn:corn}
2\mathcal{V}_\alpha(c) =& \sum_{i=1}^{n_x-1}\sum_{j=1}^{n_y-1}( S_\alpha(V^1_{i,j})  +  S_\alpha(V^2_{i,j})) \nonumber \\ 
	& -(n_y-1)\sum_{i=1}^{n_x-1}  S_\alpha(L^y_i) \,\,\,  \\ & - (n_x -1)\sum_{j=1}^{n_y-1}  S_\alpha(L^x_j), 	
\nonumber
\eeqa
where $V^1_{i,j}$ denotes a region $A$ including an $i \times j$-site rectangle of cluster $c$.  
It is more intuitive to say that we measure the four terms in Figure \ref{entanglement}(b) for each of the plaquette in \ref{entanglement}(a), which easily extends to clusters of other shapes and sizes.
In practice, due to the symmetry of the rectangular clusters all $V^2$ measurements will already be done by $V^1$.  The number of measurements required can be further reduced for a square cluster, where $V_{i,j} = V_{j,i}$ for both $V^1$ and $V^2$.

\subsection{NLCE cluster solvers}

\subsubsection{Lanczos \& Full Diagonalization}
It has not yet been discussed {\it how} the measurements on the different clusters are obtained.
In previous NLCE studies, generally the Lanczos algorithm for diagonalization was used obtain the full ground-state wavefunction of the cluster,
$\lvert \Psi_c \rangle$. 
To extract the entanglement entropy for each of the different entanglement cut geometries (see Fig.~\ref{entanglement} for examples) one must take a partial trace of the full density matrix  $\rho = \lvert \Psi_c \rangle \langle \Psi_c \rvert$ over the states in region $B$.
The partial trace is done by rewriting the ground-state vector in matrix form,
\beq
\lvert \Psi_c \rangle = \sum_i   a_i \lvert \psi^{i}_c \rangle \rightarrow M_c = \sum_{j,k} a_{jk}\lvert \psi_c^{A_j} \rangle \langle \psi_c^{B_k} \rvert,
\eeq
where $a_i$, $a_{jk}$ are numerical coefficients such that $\sum | a_i |^2 = \sum |a_{jk}|^2 = 1, \{\psi^{i}_c\}$ are basis vectors for the full cluster of both regions $A$ and $B$, and $\{\psi_c^{A_j}\}$ and  $\{\psi_c^{B_k}\}$ are the basis vectors of region $A$ and $B$ respectively.
Computationally, this means constructing a matrix with all region $A$ basis states as the rows and all region $B$ basis states are the columns.  Then, running through $\lvert \Psi_c \rangle$, each entry is assigned to an element of $M_c$ where $\psi^i_c = \psi^{A_j}_c\otimes \psi^{B_k}_c$.
From there, obtaining the reduced density matrix simply requires multiplying this matrix by its conjugate transpose,
\beqa
\rho_A &=& M_c M_c^\dagger = \sum_{i,j} \sum_{k,l}  a_{ij} a_{kl}^* \lvert \psi_c^{A_i} \rangle \langle \psi_c^{B_j} \rvert  \psi_c^{B_l} \rangle \langle \psi_c^{A_k} \rvert \nonumber \\
&=& \sum_{i,k} \sum_{j}  a_{ij} a_{kj}^* \langle \psi_c^{B_j} \rvert  \psi_c^{B_j} \rangle \lvert \psi_c^{A_i} \rangle \langle \psi_c^{A_k} \rvert.  
\eeqa
The multiplication $M_c M_c^\dagger = \rho_A$ or  $M_c^\dagger M_c = \rho_B $ is chosen based on which will result in the smaller reduced density matrix, since one final diagonalization must be done to extract its eigenvalues.
The diagonalization of the reduced density matrix must give the full eigenvalue spectrum; this requires a computationally more expensive algorithm than Lanczos which only returns the largest eigenvalues.
The above method is limited by computer memory and time, and it is suitable only for smaller clusters.  
It is imperative to use a good linear algebra library as it will significantly improve the speed and performance of this algorithm.  For this work the Eigen C++ template library\cite{eigenweb} was used to solve clusters of up to 30 sites with Lanczos and exact diagonalization.

\begin{figure}[t]
\includegraphics[width=3.2in]{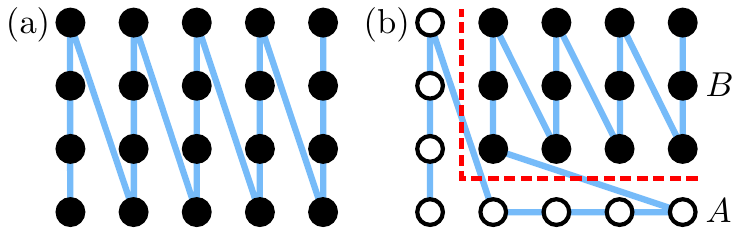}
\caption[]{
A typical path (a) used within DMRG for computing the ground-state of the 5\tim4\tim2 cluster. Here we show only
the top layer of sites; the path visits each bottom-layer site in turn before going to the next top-layer site.
An irregular path (b) is needed to obtain the corner cut dividing the system into regions $A$ and $B$ as shown.
\label{fig:dmrg}
}\end{figure}

\subsubsection{Density Matrix Renormalization Group}
For larger clusters we use a complementary method, the density matrix renormalization group (DMRG).\cite{Schollwoeck:2011} Unlike exact diagonalization and Lanczos, DMRG is not necessarily limited by the total number of sites in the cluster. 
Instead, DMRG traverses the system along a one-dimensional path, such as the one shown in Fig.~\ref{fig:dmrg}(a) for the 5\tim4\tim2 cluster (top-down view). For a path of this type, DMRG scales exponentially in $n_y$ but scales much more favorably in $n_x$ (e.g.\ linearly in $n_x$ if the system is gapped).

A key advantage of using DMRG for computing entanglement entropy is that it diagonalizes the reduced density matrix for a
different bipartition of the system at every step. These bipartitions correspond to cutting the one-dimensional path at each bond.
Thus one can obtain the full entanglement spectrum for many of the cuts needed for
the corner entanglement contribution within a single DMRG calculation. For example, the path in Fig.~\ref{fig:dmrg}(a) provides two inequivalent vertical-line
cuts (on the 4th and 8th bonds of the DMRG path) as well as corner cuts separating sites in the first column from the rest of the system. Other corner
cuts, such as the one separating the system into the regions $A$ and $B$ of Fig.~\ref{fig:dmrg}(b), require modifying the DMRG path. Such irregular
paths typically increase the computational cost needed for DMRG to reached a fixed accuracy since short-range interactions in the two-dimensional
Hamiltonian get mapped to much longer-ranged interactions in the one-dimensional model seen by DMRG.

For the results below, we used DMRG to solve the 4\tim4\tim2, 3\tim6\tim2, 6\tim3\tim2, 4\tim5\tim2, and 5\tim4\tim2 clusters. 
Each calculation kept up to $10,000$ states or enough states to obtain a truncation error below $10^{-12}$, whichever
occurred first. The only exception was one of the irregular paths for the 4\tim5\tim2 cluster for which we only managed $8000$ states (due
to memory constraints) for a truncation error of $1.4\times10^{-9}$, which is still quite accurate.
One can easily benchmark the accuracy for such difficult clusters by comparing the energy to the same cluster studied with a 
path more favorable to DMRG. For the 4\tim5\tim2 cluster with the difficult irregular path, for example, we still obtained 
the energy within a relative error of $10^{-9}$ compared to the 5\tim4\tim2 result, which is essentially exact to numerical precision.

\section{Results}
\label{sec:results}

In this section we present our results for the subleading logarithmic term in the Renyi entropy resulting from a corner of angle $\pi/2$ at the quantum critical point in the Heisenberg bilayer. We compare these to the analogous results for the quantum critical point in the $2d$ transverse-field Ising model (TFIM) on the square lattice,\cite{Kallin_NLCE} in order to study the relation between the two.

The TFIM has only a ${\mathbb Z}_2$ spin-flip symmetry, so applying the Landau-Ginzburg approach to it yields a scalar field theory with a single component. This has the same $O(N)$ $\phi^4$ action given in Eq.~(\ref{eq:action}) for the Heisenberg bilayer, except for Ising, $\vec{\phi}$ has a single component. The critical theory in this $N=1$ case is the famed Wilson-Fisher fixed point, in the same universality class of the classical 3 dimensional Ising model. This theory is not free; the NLCE calculation\cite{Kallin_NLCE} gives values of $a_\alpha(\pi/2)$ very close to, but not exactly, those calculated for Renyi index $\alpha=1,2,3$ in Gaussian free-field theory.\cite{logcorner}
This indicates that the two critical points are ``close'', in harmony with the fact that critical properties of the Wilson-Fisher fixed point can be computed very accurately in the $d=3-\epsilon$ expansion around a free-field theory.\cite{ZJ}

As discussed in Section \ref{sec:bilayer}, the Heisenberg bilayer quantum critical point is described by the 
three-component $\phi^4$ theory. Thus if $a_{\alpha}(\pi/2)$ provides a measure of the number of degrees of freedom analogous to $c$, it should be approximately three times the value obtained for the Ising theory. It need not (and ought not)  be exactly thrice the value, since this is an interacting fixed point, and critical properties depend on the number of components $N$ in a non-trivial way. Nevertheless, the $O(N)$ Wilson-Fisher fixed point can be reached by perturbing around $N$ free fields, and so one expects $a_\alpha(\pi/2)$ to be close to the free-field value, roughly $N$ times the single-component value.

\begin{figure}[t]
\includegraphics[width=3.6in]{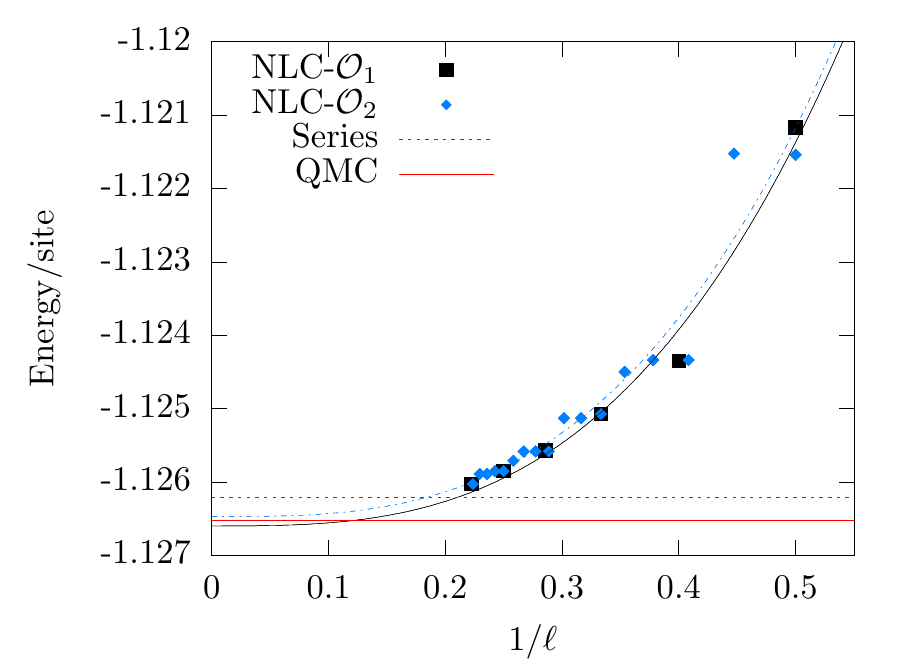}
\caption[]{
The energy per site for the Heisenberg bilayer system at its quantum critical point as a function of $1/\ell$ along with fits to $b/\ell^3 + d$ for some constants $b$ and $d$.
The horizontal lines are predictions for the ground-state energy in the thermodynamic limit from 
 series expansion and quantum Monte Carlo.\cite{AndersPC}
\label{energy}
}\end{figure}

\begin{figure}[t]
\includegraphics[width=3.7in]{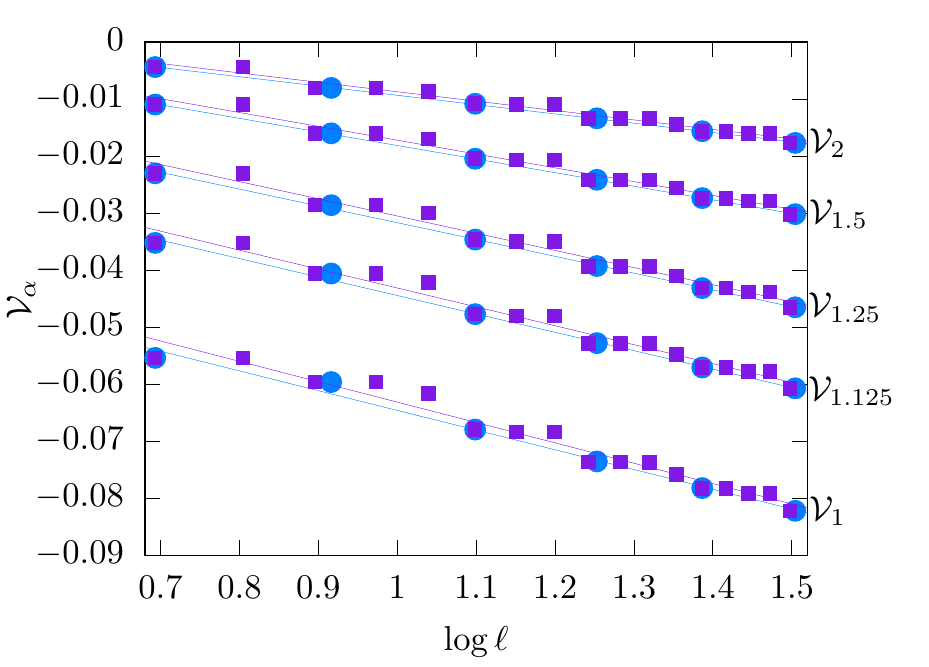}
\caption[]{
The entanglement due to a single corner $\mathcal{V}_\alpha$ in the Heisenberg bilayer system for $\alpha = 1, 1.125, 1.25, 1.5, 2$ using both definitions of order $\bigo_1$ (circles) and $\bigo_2$ (squares) along with fits to $\mathcal{V}_\alpha = a_{\alpha}\log \ell + b_\alpha$.  The resulting coefficient $a_\alpha$ is shown in figure \ref{corner} for both definitions of order.
\label{fits}
}
\end{figure}

\begin{figure}[t]
\includegraphics*[width=3.5in]{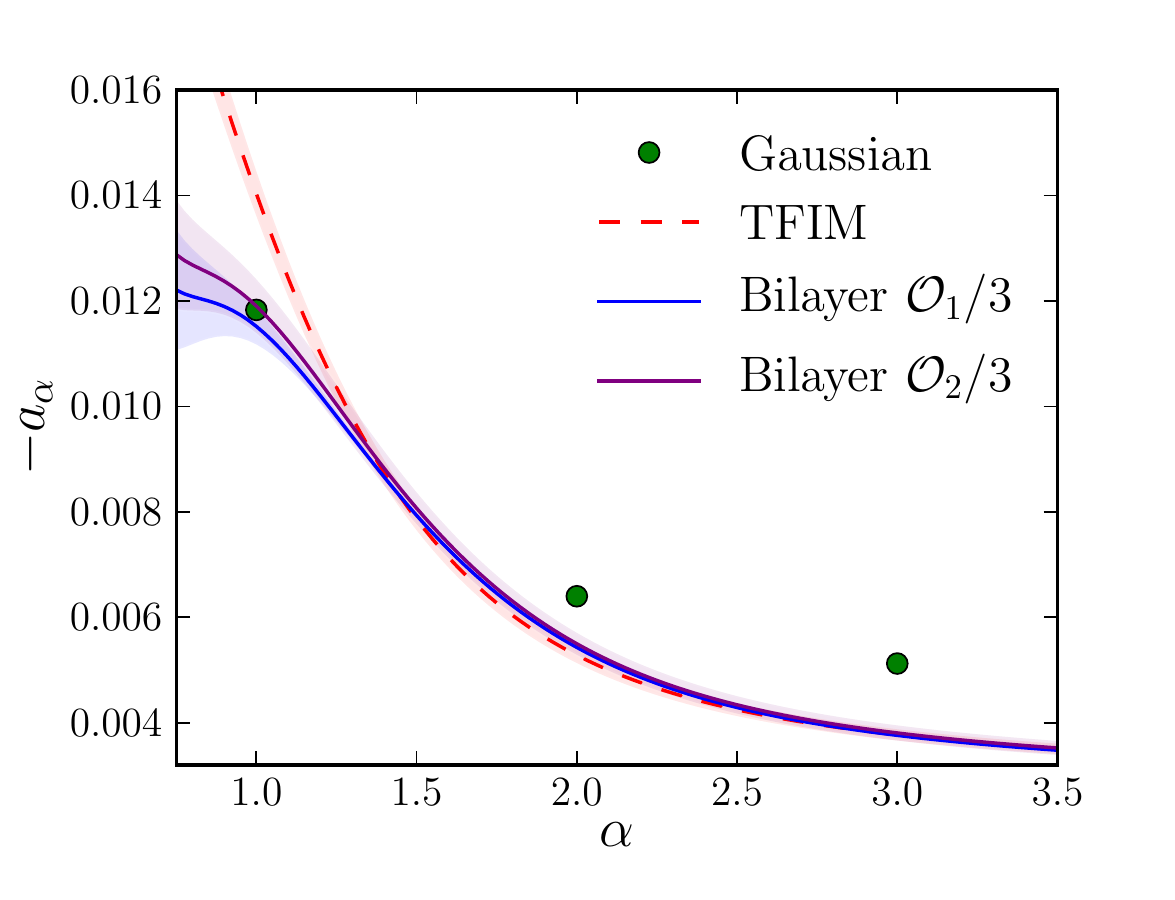}
\caption[]{The coefficient $-a_\alpha$ of the logarithm due to the presence of a $90^\circ$ corner.
Green circles are the Gaussian free field calculation.\cite{logcorner}
The red dashed line is for the transverse-field Ising model.\cite{Kallin_NLCE} 
The solid lines are results for the Heisenberg bilayer divided by 3,
with the NLCE fit using the two different definitions of the order  $\mathcal{O}_1$ and $\mathcal{O}_2$.
The shaded regions around each line correspond to an estimate of the error in the data, discussed in Section \ref{sec:results}.
\label{corner}
}
\end{figure}

As we now describe, our numerical calculations do indeed yield this factor of $3$.  We perform the NLCE procedure described in the previous section, at the critical coupling $(J_\perp/J)_c = 2.5220$ of the Heisenberg bilayer system, Eq.~(\ref{eq:ham}).
In figure \ref{energy} the NLCE includes clusters from 1\tim1\tim2 to 4\tim5\tim2, while figs.~\ref{fits} and \ref{corner} exclude the 1$d$ clusters which do not contribute to the corner entanglement term (see Eq.~\eqref{eqn:corn}).
The largest NLCE truncation orders included are $\bigo_1=4.5$ and $\bigo_2=20$.

Before discussing results for the Renyi entropies, we perform an initial check of our 
NLCE procedure, by using it to compute the ground-state energy per site in the Heisenberg bilayer at its critical point. 
At $T=0$, the ground-state energy plays the role of free energy, and by hyperscaling, its singular piece scales
as $\xi^{-(d+z)}$, where for our problem $d+z=3$. Thus, we expect the correction at order $\ell$ to scale as $1/\ell^3$.
In Fig.~\ref{energy}, the ground-state energy per site is plotted as a function of $1/\ell$ for both definitions of the order: $\bigo_1=(n_x + n_y)/2 = \ell$ and $\bigo_2 = n_x n_y = \ell^2$.  Each dataset is fit to the function $E_0(\ell) = 1/\ell^3 + E_0(\infty)$, where $E_0(\infty)$ is the predicted ground-state energy per site in the thermodynamic limit. 
For the two different fits we find,
\beqa
\bigo_1: E_0(\infty)= -1.12665 \  \nonumber\\
\bigo_2: E_0(\infty)= -1.12649  \nonumber .
\eeqa
Even though relatively small cluster sizes are included in this extrapolation, we see that both values of $E_0(\infty)$
are very close to two independent calculations from complementary, but very different techniques (see Fig.~\ref{energy}).
First, from series expansions,
Pade extrapolations lead to a value of $E_0({\infty})=-1.1262$ at $(J_\perp/J)_c = 2.5220$.
Second, from large-scale unbiased quantum Monte Carlo (QMC) Sen and Sandvik\cite{AndersPC} 
reveal a highly accurate value of $E_0({\infty})=-1.1265201(5)$ at the quantum critical point, which is again consistent with the NLCE results.


We now turn to our NLCE calculation for the subleading scaling term $\gamma$ induced by a $90^\circ$ corner in the entanglement boundary.
As discussed in the last section, the NLCE can isolate the corner contribution to the Renyi entanglement entropy independently from the 
leading ``area law'' contribution to scaling.  Thus for each value of the Renyi index $\alpha$, we extract the value of $a_\alpha(\pi/2)$ directly from fits of this corner 
entropy to
\beq
\mathcal{V}_\alpha(\ell) = a_\alpha \log \ell + b_\alpha. \label{eqn:vertex}
\eeq 
The raw NLCE data for this quantity is shown in Fig.~\ref{fits}, for several values of $\alpha$.  
Data is plotted separately for both definitions of order, $\bigo_1$ and $\bigo_2$, and separate fits are performed for each value of 
$\alpha$ to Eq.~\eqref{eqn:vertex}, as a function of the cluster length scale $\ell$, to extract $a_{\alpha}$ and $b_{\alpha}$.
We note that using $\bigo_1$ results in a systematically higher value of $b_\alpha$, though the difference decreases as $\alpha$ increases.

Using this fitting procedure, our results for the log-coefficient $a_\alpha(\pi/2)$ are plotted in Fig.~\ref{corner}. 
Here, the coefficient $-a_\alpha$ is compared between three different theories: the single-component $\phi^4$ theory via the TFIM, the free field theory calculation of Ref.~\onlinecite{logcorner}, and the present calculation of the Heisenberg bilayer. 
In Fig.~\ref{corner}, data for the Heisenberg bilayer is divided by 3 to emphasize how remarkably close it is to thrice the Ising data. 
The implications of this are discussed in detail in section \ref{sec:conclusion}.

The error bars shown in Fig.~\ref{corner} are meant to be a guide to the reader.
They are calculated as the standard deviation of the data from the linear fits to $\mathcal{V}_\alpha(\ell) = a_\alpha \log \ell + b_\alpha$, examples of which are shown in Fig.~\ref{fits}.
This error is then assigned to $a_\alpha$, although strictly speaking it also depends on $b_\alpha$.
As with any study of this kind which incorporates functional extrapolations using relatively small cluster sizes, significant uncertainty 
related to the precise data series included in the fit remains, and is not represented by the error bars in Fig.~\ref{corner}.
It is worth noting that the NLCE results\cite{Kallin_NLCE} for the second Renyi entropy $S_2(A)$ in the TFIM were independently benchmarked against series-expansion\cite{TFIM_series}
and QMC data\cite{Tommaso,Inglis_QMC} obtained through a replica-trick procedure.  Both calculations 
yield a coefficient $a(\pi/2)$ consistent with the NLCE to within numerical errors.
The unbiased QMC data were obtained from a much different fitting procedure involving a square subregion
$A$ with four corners embedded in a toroidal lattice; thus the match with the NLCE is particularly striking.

It is clear from Fig.~\ref{corner} that some uncertainty remains regarding the relationship between the data for 
$\alpha \lesssim 1.3$, obfuscated by the growth in uncertainties in the NLCE calculation in this regime.
One can see from Fig.~\ref{fits} that the data points deviate further above and below the linear fit for smaller values of $\alpha$,
directly increasing the error bars.
The poorer numerical convergence of the NLCE for smaller $\alpha$ could be expected, particularly 
since the $\alpha < 1$ Renyi entropies are more sensitive to the tail of the entanglement spectrum. 
At least for gapped systems, it has been argued that the closer a reduced density matrix eigenstate's 
eigenvalue is to the tail of the entanglement spectrum, the further it probes the system away from the cut.\cite{Ari}
Thus, although it seems likely that the factor of 3 relationship between the TFIM and Heisenberg bilayer data 
remains for $\alpha \lesssim 1.3$, 
it remains possible that deviations, corrections, or even a phase transition in $\alpha$ change the relationship.\cite{Max}

Despite the uncertainties, Fig.~\ref{corner} gives substantial support to the hypothesis that $a_\alpha(\pi/2)$ for the Heisenberg bilayer at its critical point
is approximately thrice the value for the TFIM at its critical point.

\section{Summary And Discussion} 
\label{sec:conclusion}

We have studied the Renyi entanglement entropies for a square-lattice Heisenberg bilayer antiferromagnet with a Numerical Linked Cluster Expansion method,
employing both Lanczos and DMRG techniques as cluster solvers.
Focussing on the subleading logarithmic scaling contribution to the Renyi entropies that arises from a 90$^{\circ}$ corner,
we calculate the cutoff-independent coefficient $a_{\alpha}(\pi/2)$ of the logarithm directly at the quantum critical point of the model.
This critical point is in the $3D$ $O(3$) universality class,
described by a three-component $\phi^4$ field theory.
We find that the universal coefficient is thrice that computed previously\cite{Kallin_NLCE} for a quantum critical point in the $3D$ Ising universality class,
described by a single-component $\phi^4$ field theory. Our calculation thus provides substantial evidence that this universal coefficient provides a measure of the number of
degrees of freedom of the field theory. In this interacting but close-to-free theory, this is the number of
low-lying bosonic modes present at the quantum critical point.

There have been hints in past literature that the coefficient of the corner-induced logarithm contains valuable information about the effective low-energy critical theory.
Most concretely, for $z\ne 1$ conformal quantum field theories, the coefficient is proportional to the central charge of the conformal field theory used to build the ground-state wavefunction.\cite{FradkinMoore} Since we found analogous behavior in $z=1$ (Lorentz-invariant) field theories,
it is thus tantalizing to speculate that this coefficient provides a quantity analogous to a central charge for any  $2+1$-dimensional quantum critical point. 

Holographic calculations of the entanglement entropy in higher dimensions also suggest this behavior.
Using the AdS/CFT correspondence, the corner-induced logarithm from an angle $\theta$ and its analogs in arbitrary dimensions obey the scaling form \cite{Robcorner}
\begin{equation}
\gamma = \left(\frac{L}{L_P}\right)^d q(\theta) \log \left({ \frac{\ell}{\delta} }\right)\ +\ \dots,
\end{equation}
where $L$ is the length scale set by the AdS curvature and $L_P$ is the Planck length, while the ellipses include a $\log^2(\ell/\delta)$ term for odd $d>1$.  The Planck length arises from Newton's constant, which appears in Ryu and Takayangi's famous formula relating the entanglement entropy to an area of a minimal surface in the AdS space.\cite{ryu} The factor $(L/L_P)^d$ scales appropriately for counting the degrees of freedom in $d$-dimensional space with the Planck length as a short-distance cutoff, and the cutoff-independent function $q(\theta)$ has been computed for $d<6$. In $d=1$, the coefficient indeed becomes precisely $c/3$,\cite{Brown} and in $d>2$ other relations with conformal central charges can be found.\cite{Robcorner} Thus it is natural to believe that the coefficient of the log term in our case $d=2$ is also proportional to some universal quantity like a central charge giving a measure for the degrees of freedom. Our results provide support for this idea.

Finally, it is important to test this behavior on other $2+1$ critical points with different types of effective low-energy theories.
From this work, the conjecture is that the coefficient of the corner-induced logarithm counts a very clear signal (i.e.~the integer $N$ in the $O(N)$ theory), and hence may be relatively insensitive to some finite-size effects.
These finite-size effects will continue to be reduced in the future with the further adoption in NLCE of the 
powerful and general DMRG method, which has been shown to provide
very accurate results on quasi-$2d$ finite-size systems.\cite{Stoudenmire}
Thus, the calculations in this paper can immediately and straightforwardly be extended to other lattice models.

An obvious next candidate for study is an $O(2)$ critical point in $2+1d$, such as occurs in a spin 1/2 XY model with bilayer couplings,
or alternatively, in a symmetry-breaking magnetic field.
If these and other conventional critical points confirm our scenario, then
especially important would be the study of critical points with exotic structures to their low-energy effective theories.
An exciting prospect would be the $SU(2)$-invariant Heisenberg model with four-site exchange (Sandvik's J-Q model\cite{JQ}), which is believed to 
contain a critical point in the non-compact CP$^1$ universality class, a field theory that describes two flavors of spinons interacting with a non-compact $U(1)$ gauge field.\cite{DQCP,ARCMP_Kaul}

\subsection*{Acknowledgments} 
We would like to acknowledge crucial discussions with A. Burkov, A. John Berlinsky, J. Carrasquilla, H. Casini, A. Ferris, S. Inglis, R. Myers, W. Witczak-Krempa, and A. Sandvik.
The simulations were performed on the computing facilities of SHARCNET and on the Perimeter Institute HPC.  Support was provided by NSERC, the Canada Research Chair program, the Ontario Ministry of Research and Innovation, the John Templeton Foundation, and the Perimeter Institute (PI) for Theoretical Physics. Research at PI is supported by the Government of Canada through Industry Canada and by the Province of Ontario through the Ministry of Economic Development \& Innovation.  PF is supported by National Science Foundation grant DMR/MPS1006549 and RRPS is supported by NSF grant DMR-1004231.

\bibliography{Biblio}

\begin{thebibliography}{59}%
\makeatletter
\providecommand \@ifxundefined [1]{%
 \@ifx{#1\undefined}
}%
\providecommand \@ifnum [1]{%
 \ifnum #1\expandafter \@firstoftwo
 \else \expandafter \@secondoftwo
 \fi
}%
\providecommand \@ifx [1]{%
 \ifx #1\expandafter \@firstoftwo
 \else \expandafter \@secondoftwo
 \fi
}%
\providecommand \natexlab [1]{#1}%
\providecommand \enquote  [1]{``#1''}%
\providecommand \bibnamefont  [1]{#1}%
\providecommand \bibfnamefont [1]{#1}%
\providecommand \citenamefont [1]{#1}%
\providecommand \href@noop [0]{\@secondoftwo}%
\providecommand \href [0]{\begingroup \@sanitize@url \@href}%
\providecommand \@href[1]{\@@startlink{#1}\@@href}%
\providecommand \@@href[1]{\endgroup#1\@@endlink}%
\providecommand \@sanitize@url [0]{\catcode `\\12\catcode `\$12\catcode
  `\&12\catcode `\#12\catcode `\^12\catcode `\_12\catcode `\%12\relax}%
\providecommand \@@startlink[1]{}%
\providecommand \@@endlink[0]{}%
\providecommand \url  [0]{\begingroup\@sanitize@url \@url }%
\providecommand \@url [1]{\endgroup\@href {#1}{\urlprefix }}%
\providecommand \urlprefix  [0]{URL }%
\providecommand \Eprint [0]{\href }%
\providecommand \doibase [0]{http://dx.doi.org/}%
\providecommand \selectlanguage [0]{\@gobble}%
\providecommand \bibinfo  [0]{\@secondoftwo}%
\providecommand \bibfield  [0]{\@secondoftwo}%
\providecommand \translation [1]{[#1]}%
\providecommand \BibitemOpen [0]{}%
\providecommand \bibitemStop [0]{}%
\providecommand \bibitemNoStop [0]{.\EOS\space}%
\providecommand \EOS [0]{\spacefactor3000\relax}%
\providecommand \BibitemShut  [1]{\csname bibitem#1\endcsname}%
\let\auto@bib@innerbib\@empty
\bibitem [{EEr(2009)}]{EEreview}%
  \BibitemOpen
  \href@noop {} {\emph {\bibinfo {title} {{Entanglement entropy in extended
  quantum systems}}}},\ \bibinfo {series} {{J. Phys. A (special issue)}}, Vol.\
  \bibinfo {volume} {424}\ (\bibinfo {year} {2009})\BibitemShut {NoStop}%
\bibitem [{\citenamefont {Casini}\ and\ \citenamefont
  {Huerta}(2012)}]{Casini12}%
  \BibitemOpen
  \bibfield  {author} {\bibinfo {author} {\bibfnamefont {H.}~\bibnamefont
  {Casini}}\ and\ \bibinfo {author} {\bibfnamefont {M.}~\bibnamefont
  {Huerta}},\ }\href {\doibase 10.1103/PhysRevD.85.125016} {\bibfield
  {journal} {\bibinfo  {journal} {Phys.Rev.}\ }\textbf {\bibinfo {volume}
  {D85}},\ \bibinfo {pages} {125016} (\bibinfo {year} {2012})},\ \Eprint
  {http://arxiv.org/abs/1202.5650} {arXiv:1202.5650 [hep-th]} \BibitemShut
  {NoStop}%
\bibitem [{\citenamefont {Grover}(2012)}]{Grover_C}%
  \BibitemOpen
  \bibfield  {author} {\bibinfo {author} {\bibfnamefont {T.}~\bibnamefont
  {Grover}},\ }\href@noop {} {\  (\bibinfo {year} {2012})},\ \Eprint
  {http://arxiv.org/abs/arXiv:1211.1392} {arXiv:1211.1392} \BibitemShut
  {NoStop}%
\bibitem [{\citenamefont {Zamolodchikov}(1986)}]{Zamo}%
  \BibitemOpen
  \bibfield  {author} {\bibinfo {author} {\bibfnamefont {A.}~\bibnamefont
  {Zamolodchikov}},\ }\href@noop {} {\bibfield  {journal} {\bibinfo  {journal}
  {JETP Lett.}\ }\textbf {\bibinfo {volume} {43}},\ \bibinfo {pages} {731}
  (\bibinfo {year} {1986})}\BibitemShut {NoStop}%
\bibitem [{\citenamefont {Holzhey}\ \emph {et~al.}(1994)\citenamefont
  {Holzhey}, \citenamefont {Larsen},\ and\ \citenamefont {Wilczek}}]{Holzhey}%
  \BibitemOpen
  \bibfield  {author} {\bibinfo {author} {\bibfnamefont {C.}~\bibnamefont
  {Holzhey}}, \bibinfo {author} {\bibfnamefont {F.}~\bibnamefont {Larsen}}, \
  and\ \bibinfo {author} {\bibfnamefont {F.}~\bibnamefont {Wilczek}},\ }\href
  {\doibase 10.1016/0550-3213(94)90402-2} {\bibfield  {journal} {\bibinfo
  {journal} {Nucl.Phys. B}\ }\textbf {\bibinfo {volume} {424}},\ \bibinfo
  {pages} {443} (\bibinfo {year} {1994})},\ \Eprint
  {http://arxiv.org/abs/hep-th/9403108} {arXiv:hep-th/9403108 [hep-th]}
  \BibitemShut {NoStop}%
\bibitem [{\citenamefont {Vidal}\ \emph {et~al.}(2003)\citenamefont {Vidal},
  \citenamefont {Latorre}, \citenamefont {Rico},\ and\ \citenamefont
  {Kitaev}}]{VLRK}%
  \BibitemOpen
  \bibfield  {author} {\bibinfo {author} {\bibfnamefont {G.}~\bibnamefont
  {Vidal}}, \bibinfo {author} {\bibfnamefont {J.~I.}\ \bibnamefont {Latorre}},
  \bibinfo {author} {\bibfnamefont {E.}~\bibnamefont {Rico}}, \ and\ \bibinfo
  {author} {\bibfnamefont {A.}~\bibnamefont {Kitaev}},\ }\href {\doibase
  10.1103/PhysRevLett.90.227902} {\bibfield  {journal} {\bibinfo  {journal}
  {Phys.Rev.Lett.}\ }\textbf {\bibinfo {volume} {90}},\ \bibinfo {pages}
  {227902} (\bibinfo {year} {2003})},\ \Eprint
  {http://arxiv.org/abs/quant-ph/0211074} {arXiv:quant-ph/0211074 [quant-ph]}
  \BibitemShut {NoStop}%
\bibitem [{\citenamefont {Korepin}(2004)}]{Korepin}%
  \BibitemOpen
  \bibfield  {author} {\bibinfo {author} {\bibfnamefont {V.~E.}\ \bibnamefont
  {Korepin}},\ }\href@noop {} {\bibfield  {journal} {\bibinfo  {journal} {Phys.
  Rev. Lett.}\ }\textbf {\bibinfo {volume} {92}},\ \bibinfo {pages} {096402}
  (\bibinfo {year} {2004})}\BibitemShut {NoStop}%
\bibitem [{\citenamefont {Calabrese}\ and\ \citenamefont
  {Cardy}(2004)}]{Cardy}%
  \BibitemOpen
  \bibfield  {author} {\bibinfo {author} {\bibfnamefont {P.}~\bibnamefont
  {Calabrese}}\ and\ \bibinfo {author} {\bibfnamefont {J.}~\bibnamefont
  {Cardy}},\ }\href@noop {} {\bibfield  {journal} {\bibinfo  {journal} {J.
  Stat. Mech.: Theor. Exp.}\ }\textbf {\bibinfo {volume} {P06002}} (\bibinfo
  {year} {2004})}\BibitemShut {NoStop}%
\bibitem [{\citenamefont {Bombelli}\ \emph {et~al.}(1986)\citenamefont
  {Bombelli}, \citenamefont {Koul}, \citenamefont {Lee},\ and\ \citenamefont
  {Sorkin}}]{Sorkin}%
  \BibitemOpen
  \bibfield  {author} {\bibinfo {author} {\bibfnamefont {L.}~\bibnamefont
  {Bombelli}}, \bibinfo {author} {\bibfnamefont {R.~K.}\ \bibnamefont {Koul}},
  \bibinfo {author} {\bibfnamefont {J.}~\bibnamefont {Lee}}, \ and\ \bibinfo
  {author} {\bibfnamefont {R.~D.}\ \bibnamefont {Sorkin}},\ }\href {\doibase
  10.1103/PhysRevD.34.373} {\bibfield  {journal} {\bibinfo  {journal} {Phys.
  Rev. D}\ }\textbf {\bibinfo {volume} {34}},\ \bibinfo {pages} {373} (\bibinfo
  {year} {1986})}\BibitemShut {NoStop}%
\bibitem [{\citenamefont {Srednicki}(1993)}]{Shredder}%
  \BibitemOpen
  \bibfield  {author} {\bibinfo {author} {\bibfnamefont {M.}~\bibnamefont
  {Srednicki}},\ }\href {\doibase 10.1103/PhysRevLett.71.666} {\bibfield
  {journal} {\bibinfo  {journal} {Phys. Rev. Lett.}\ }\textbf {\bibinfo
  {volume} {71}},\ \bibinfo {pages} {666} (\bibinfo {year} {1993})}\BibitemShut
  {NoStop}%
\bibitem [{\citenamefont {Eisert}\ \emph {et~al.}(2010)\citenamefont {Eisert},
  \citenamefont {Cramer},\ and\ \citenamefont {Plenio}}]{ALreview}%
  \BibitemOpen
  \bibfield  {author} {\bibinfo {author} {\bibfnamefont {J.}~\bibnamefont
  {Eisert}}, \bibinfo {author} {\bibfnamefont {M.}~\bibnamefont {Cramer}}, \
  and\ \bibinfo {author} {\bibfnamefont {M.~B.}\ \bibnamefont {Plenio}},\
  }\href {\doibase 10.1103/RevModPhys.82.277} {\bibfield  {journal} {\bibinfo
  {journal} {Rev. Mod. Phys.}\ }\textbf {\bibinfo {volume} {82}},\ \bibinfo
  {pages} {277} (\bibinfo {year} {2010})}\BibitemShut {NoStop}%
\bibitem [{\citenamefont {Wolf}(2006)}]{Wolf2}%
  \BibitemOpen
  \bibfield  {author} {\bibinfo {author} {\bibfnamefont {M.~M.}\ \bibnamefont
  {Wolf}},\ }\href {\doibase 10.1103/PhysRevLett.96.010404} {\bibfield
  {journal} {\bibinfo  {journal} {Phys. Rev. Lett.}\ }\textbf {\bibinfo
  {volume} {96}},\ \bibinfo {pages} {010404} (\bibinfo {year}
  {2006})}\BibitemShut {NoStop}%
\bibitem [{\citenamefont {Gioev}\ and\ \citenamefont {Klich}(2006)}]{Israel}%
  \BibitemOpen
  \bibfield  {author} {\bibinfo {author} {\bibfnamefont {D.}~\bibnamefont
  {Gioev}}\ and\ \bibinfo {author} {\bibfnamefont {I.}~\bibnamefont {Klich}},\
  }\href {\doibase 10.1103/PhysRevLett.96.100503} {\bibfield  {journal}
  {\bibinfo  {journal} {Phys. Rev. Lett.}\ }\textbf {\bibinfo {volume} {96}},\
  \bibinfo {pages} {100503} (\bibinfo {year} {2006})}\BibitemShut {NoStop}%
\bibitem [{\citenamefont {Lai}\ \emph {et~al.}(2013)\citenamefont {Lai},
  \citenamefont {Yang},\ and\ \citenamefont {Bonesteel}}]{EBL}%
  \BibitemOpen
  \bibfield  {author} {\bibinfo {author} {\bibfnamefont {H.-H.}\ \bibnamefont
  {Lai}}, \bibinfo {author} {\bibfnamefont {K.}~\bibnamefont {Yang}}, \ and\
  \bibinfo {author} {\bibfnamefont {N.~E.}\ \bibnamefont {Bonesteel}},\ }\href
  {\doibase 10.1103/PhysRevLett.111.210402} {\bibfield  {journal} {\bibinfo
  {journal} {Phys. Rev. Lett.}\ }\textbf {\bibinfo {volume} {111}},\ \bibinfo
  {pages} {210402} (\bibinfo {year} {2013})}\BibitemShut {NoStop}%
\bibitem [{\citenamefont {Fursaev}(2006)}]{Fursaev}%
  \BibitemOpen
  \bibfield  {author} {\bibinfo {author} {\bibfnamefont {D.~V.}\ \bibnamefont
  {Fursaev}},\ }\href {\doibase 10.1103/PhysRevD.73.124025} {\bibfield
  {journal} {\bibinfo  {journal} {Phys. Rev.}\ }\textbf {\bibinfo {volume}
  {D73}},\ \bibinfo {pages} {124025} (\bibinfo {year} {2006})},\ \Eprint
  {http://arxiv.org/abs/hep-th/0602134} {arXiv:hep-th/0602134 [hep-th]}
  \BibitemShut {NoStop}%
\bibitem [{\citenamefont {Metlitski}\ \emph {et~al.}(2009)\citenamefont
  {Metlitski}, \citenamefont {Fuertes},\ and\ \citenamefont {Sachdev}}]{Max}%
  \BibitemOpen
  \bibfield  {author} {\bibinfo {author} {\bibfnamefont {M.~A.}\ \bibnamefont
  {Metlitski}}, \bibinfo {author} {\bibfnamefont {C.~A.}\ \bibnamefont
  {Fuertes}}, \ and\ \bibinfo {author} {\bibfnamefont {S.}~\bibnamefont
  {Sachdev}},\ }\href@noop {} {\bibfield  {journal} {\bibinfo  {journal} {Phys.
  Rev. B}\ }\textbf {\bibinfo {volume} {80}},\ \bibinfo {pages} {115122}
  (\bibinfo {year} {2009})}\BibitemShut {NoStop}%
\bibitem [{Note1()}]{Note1}%
  \BibitemOpen
  \bibinfo {note} {For a simple explanation of how $c$ measures degrees of
  freedom in 1+1 dimensions, see Ref. 59.}\BibitemShut {Stop}%
\bibitem [{\citenamefont {Fradkin}(2013)}]{Fradkinbook}%
  \BibitemOpen
  \bibfield  {author} {\bibinfo {author} {\bibfnamefont {E.}~\bibnamefont
  {Fradkin}},\ }\href@noop {} {\emph {\bibinfo {title} {{Field Theories of
  Condensed Matter Systems}}}},\ \bibinfo {edition} {2nd}\ ed.\ (\bibinfo
  {publisher} {Cambridge University Press},\ \bibinfo {year}
  {2013})\BibitemShut {NoStop}%
\bibitem [{\citenamefont {Ryu}\ and\ \citenamefont {Takayanagi}(2006)}]{ryu}%
  \BibitemOpen
  \bibfield  {author} {\bibinfo {author} {\bibfnamefont {S.}~\bibnamefont
  {Ryu}}\ and\ \bibinfo {author} {\bibfnamefont {T.}~\bibnamefont
  {Takayanagi}},\ }\href {\doibase 10.1103/PhysRevLett.96.181602} {\bibfield
  {journal} {\bibinfo  {journal} {Phys. Rev. Lett.}\ }\textbf {\bibinfo
  {volume} {96}},\ \bibinfo {pages} {181602} (\bibinfo {year}
  {2006})}\BibitemShut {NoStop}%
\bibitem [{\citenamefont {Nishioka}\ \emph {et~al.}(2009)\citenamefont
  {Nishioka}, \citenamefont {Ryu},\ and\ \citenamefont {Takayanagi}}]{ryu_2}%
  \BibitemOpen
  \bibfield  {author} {\bibinfo {author} {\bibfnamefont {T.}~\bibnamefont
  {Nishioka}}, \bibinfo {author} {\bibfnamefont {S.}~\bibnamefont {Ryu}}, \
  and\ \bibinfo {author} {\bibfnamefont {T.}~\bibnamefont {Takayanagi}},\
  }\href {http://stacks.iop.org/1751-8121/42/i=50/a=504008} {\bibfield
  {journal} {\bibinfo  {journal} {J. Phys. A}\ }\textbf {\bibinfo {volume}
  {42}},\ \bibinfo {pages} {504008} (\bibinfo {year} {2009})}\BibitemShut
  {NoStop}%
\bibitem [{\citenamefont {Ardonne}\ \emph {et~al.}(2004)\citenamefont
  {Ardonne}, \citenamefont {Fendley},\ and\ \citenamefont {Fradkin}}]{AFF}%
  \BibitemOpen
  \bibfield  {author} {\bibinfo {author} {\bibfnamefont {E.}~\bibnamefont
  {Ardonne}}, \bibinfo {author} {\bibfnamefont {P.}~\bibnamefont {Fendley}}, \
  and\ \bibinfo {author} {\bibfnamefont {E.}~\bibnamefont {Fradkin}},\ }\href
  {\doibase 10.1016/j.aop.2004.01.004} {\bibfield  {journal} {\bibinfo
  {journal} {Annals Phys.}\ }\textbf {\bibinfo {volume} {310}},\ \bibinfo
  {pages} {493} (\bibinfo {year} {2004})},\ \Eprint
  {http://arxiv.org/abs/cond-mat/0311466} {arXiv:cond-mat/0311466 [cond-mat]}
  \BibitemShut {NoStop}%
\bibitem [{\citenamefont {Rokhsar}\ and\ \citenamefont {Kivelson}(1988)}]{RK}%
  \BibitemOpen
  \bibfield  {author} {\bibinfo {author} {\bibfnamefont {D.~S.}\ \bibnamefont
  {Rokhsar}}\ and\ \bibinfo {author} {\bibfnamefont {S.~A.}\ \bibnamefont
  {Kivelson}},\ }\href {\doibase 10.1103/PhysRevLett.61.2376} {\bibfield
  {journal} {\bibinfo  {journal} {Phys.Rev.Lett.}\ }\textbf {\bibinfo {volume}
  {61}},\ \bibinfo {pages} {2376} (\bibinfo {year} {1988})}\BibitemShut
  {NoStop}%
\bibitem [{\citenamefont {Cardy}\ and\ \citenamefont
  {Peschel}(1988)}]{cardy-peschel}%
  \BibitemOpen
  \bibfield  {author} {\bibinfo {author} {\bibfnamefont {J.~L.}\ \bibnamefont
  {Cardy}}\ and\ \bibinfo {author} {\bibfnamefont {I.}~\bibnamefont
  {Peschel}},\ }\href@noop {} {\bibfield  {journal} {\bibinfo  {journal}
  {Nuclear Physics B}\ }\textbf {\bibinfo {volume} {300}},\ \bibinfo {pages}
  {377} (\bibinfo {year} {1988})}\BibitemShut {NoStop}%
\bibitem [{\citenamefont {Fradkin}\ and\ \citenamefont
  {Moore}(2006)}]{FradkinMoore}%
  \BibitemOpen
  \bibfield  {author} {\bibinfo {author} {\bibfnamefont {E.}~\bibnamefont
  {Fradkin}}\ and\ \bibinfo {author} {\bibfnamefont {J.~E.}\ \bibnamefont
  {Moore}},\ }\href {\doibase 10.1103/PhysRevLett.97.050404} {\bibfield
  {journal} {\bibinfo  {journal} {Phys. Rev. Lett.}\ }\textbf {\bibinfo
  {volume} {97}},\ \bibinfo {pages} {050404} (\bibinfo {year}
  {2006})}\BibitemShut {NoStop}%
\bibitem [{\citenamefont {Casini}\ and\ \citenamefont
  {Huerta}(2007)}]{logcorner}%
  \BibitemOpen
  \bibfield  {author} {\bibinfo {author} {\bibfnamefont {H.}~\bibnamefont
  {Casini}}\ and\ \bibinfo {author} {\bibfnamefont {M.}~\bibnamefont
  {Huerta}},\ }\href@noop {} {\bibfield  {journal} {\bibinfo  {journal} {Nucl.
  Phys. B}\ }\textbf {\bibinfo {volume} {764}},\ \bibinfo {pages} {183}
  (\bibinfo {year} {2007})}\BibitemShut {NoStop}%
\bibitem [{\citenamefont {Kallin}\ \emph {et~al.}(2011)\citenamefont {Kallin},
  \citenamefont {Hastings}, \citenamefont {Melko},\ and\ \citenamefont
  {Singh}}]{Kallin_Heis}%
  \BibitemOpen
  \bibfield  {author} {\bibinfo {author} {\bibfnamefont {A.~B.}\ \bibnamefont
  {Kallin}}, \bibinfo {author} {\bibfnamefont {M.~B.}\ \bibnamefont
  {Hastings}}, \bibinfo {author} {\bibfnamefont {R.~G.}\ \bibnamefont {Melko}},
  \ and\ \bibinfo {author} {\bibfnamefont {R.~R.~P.}\ \bibnamefont {Singh}},\
  }\href {\doibase 10.1103/PhysRevB.84.165134} {\bibfield  {journal} {\bibinfo
  {journal} {Phys. Rev. B}\ }\textbf {\bibinfo {volume} {84}},\ \bibinfo
  {pages} {165134} (\bibinfo {year} {2011})}\BibitemShut {NoStop}%
\bibitem [{\citenamefont {Singh}\ \emph {et~al.}(2012)\citenamefont {Singh},
  \citenamefont {Melko},\ and\ \citenamefont {Oitmaa}}]{TFIM_series}%
  \BibitemOpen
  \bibfield  {author} {\bibinfo {author} {\bibfnamefont {R.~R.~P.}\
  \bibnamefont {Singh}}, \bibinfo {author} {\bibfnamefont {R.~G.}\ \bibnamefont
  {Melko}}, \ and\ \bibinfo {author} {\bibfnamefont {J.}~\bibnamefont
  {Oitmaa}},\ }\href@noop {} {\bibfield  {journal} {\bibinfo  {journal} {Phys.
  Rev. B}\ }\textbf {\bibinfo {volume} {86}},\ \bibinfo {pages} {075106}
  (\bibinfo {year} {2012})}\BibitemShut {NoStop}%
\bibitem [{\citenamefont {Kallin}\ \emph {et~al.}(2013)\citenamefont {Kallin},
  \citenamefont {Hyatt}, \citenamefont {Singh},\ and\ \citenamefont
  {Melko}}]{Kallin_NLCE}%
  \BibitemOpen
  \bibfield  {author} {\bibinfo {author} {\bibfnamefont {A.~B.}\ \bibnamefont
  {Kallin}}, \bibinfo {author} {\bibfnamefont {K.}~\bibnamefont {Hyatt}},
  \bibinfo {author} {\bibfnamefont {R.~R.~P.}\ \bibnamefont {Singh}}, \ and\
  \bibinfo {author} {\bibfnamefont {R.~G.}\ \bibnamefont {Melko}},\ }\href
  {\doibase 10.1103/PhysRevLett.110.135702} {\bibfield  {journal} {\bibinfo
  {journal} {Phys. Rev. Lett.}\ }\textbf {\bibinfo {volume} {110}},\ \bibinfo
  {pages} {135702} (\bibinfo {year} {2013})}\BibitemShut {NoStop}%
\bibitem [{\citenamefont {Inglis}\ and\ \citenamefont
  {Melko}(2013)}]{Inglis_QMC}%
  \BibitemOpen
  \bibfield  {author} {\bibinfo {author} {\bibfnamefont {S.}~\bibnamefont
  {Inglis}}\ and\ \bibinfo {author} {\bibfnamefont {R.~G.}\ \bibnamefont
  {Melko}},\ }\href {http://stacks.iop.org/1367-2630/15/i=7/a=073048}
  {\bibfield  {journal} {\bibinfo  {journal} {New Journal of Physics}\ }\textbf
  {\bibinfo {volume} {15}},\ \bibinfo {pages} {073048} (\bibinfo {year}
  {2013})}\BibitemShut {NoStop}%
\bibitem [{\citenamefont {Hirata}\ and\ \citenamefont
  {Takayanagi}(2007)}]{Hirata}%
  \BibitemOpen
  \bibfield  {author} {\bibinfo {author} {\bibfnamefont {T.}~\bibnamefont
  {Hirata}}\ and\ \bibinfo {author} {\bibfnamefont {T.}~\bibnamefont
  {Takayanagi}},\ }\href {\doibase 10.1088/1126-6708/2007/02/042} {\bibfield
  {journal} {\bibinfo  {journal} {JHEP}\ }\textbf {\bibinfo {volume} {0702}},\
  \bibinfo {pages} {042} (\bibinfo {year} {2007})},\ \Eprint
  {http://arxiv.org/abs/hep-th/0608213} {arXiv:hep-th/0608213 [hep-th]}
  \BibitemShut {NoStop}%
\bibitem [{\citenamefont {Myers}\ and\ \citenamefont
  {Singh}(2012)}]{Robcorner}%
  \BibitemOpen
  \bibfield  {author} {\bibinfo {author} {\bibfnamefont {R.~C.}\ \bibnamefont
  {Myers}}\ and\ \bibinfo {author} {\bibfnamefont {A.}~\bibnamefont {Singh}},\
  }\href@noop {} {\  (\bibinfo {year} {2012})},\ \Eprint
  {http://arxiv.org/abs/arXiv:1206.5225} {arXiv:1206.5225} \BibitemShut
  {NoStop}%
\bibitem [{\citenamefont {Wang}\ \emph {et~al.}(2006)\citenamefont {Wang},
  \citenamefont {Beach},\ and\ \citenamefont {Sandvik}}]{Wang_bilayer}%
  \BibitemOpen
  \bibfield  {author} {\bibinfo {author} {\bibfnamefont {L.}~\bibnamefont
  {Wang}}, \bibinfo {author} {\bibfnamefont {K.~S.~D.}\ \bibnamefont {Beach}},
  \ and\ \bibinfo {author} {\bibfnamefont {A.~W.}\ \bibnamefont {Sandvik}},\
  }\href {\doibase 10.1103/PhysRevB.73.014431} {\bibfield  {journal} {\bibinfo
  {journal} {Phys. Rev. B}\ }\textbf {\bibinfo {volume} {73}},\ \bibinfo
  {pages} {014431} (\bibinfo {year} {2006})}\BibitemShut {NoStop}%
\bibitem [{\citenamefont {Holm}\ and\ \citenamefont {Janke}(1993)}]{Janke}%
  \BibitemOpen
  \bibfield  {author} {\bibinfo {author} {\bibfnamefont {C.}~\bibnamefont
  {Holm}}\ and\ \bibinfo {author} {\bibfnamefont {W.}~\bibnamefont {Janke}},\
  }\href {\doibase 10.1103/PhysRevB.48.936} {\bibfield  {journal} {\bibinfo
  {journal} {Phys. Rev. B}\ }\textbf {\bibinfo {volume} {48}},\ \bibinfo
  {pages} {936} (\bibinfo {year} {1993})}\BibitemShut {NoStop}%
\bibitem [{\citenamefont {Rigol}\ \emph {et~al.}(2006)\citenamefont {Rigol},
  \citenamefont {Bryant},\ and\ \citenamefont {Singh}}]{NLC1}%
  \BibitemOpen
  \bibfield  {author} {\bibinfo {author} {\bibfnamefont {M.}~\bibnamefont
  {Rigol}}, \bibinfo {author} {\bibfnamefont {T.}~\bibnamefont {Bryant}}, \
  and\ \bibinfo {author} {\bibfnamefont {R.~R.~P.}\ \bibnamefont {Singh}},\
  }\href@noop {} {\bibfield  {journal} {\bibinfo  {journal} {Phys. Rev. Lett.}\
  }\textbf {\bibinfo {volume} {97}},\ \bibinfo {pages} {187202} (\bibinfo
  {year} {2006})}\BibitemShut {NoStop}%
\bibitem [{\citenamefont {Rigol}\ \emph
  {et~al.}(2007{\natexlab{a}})\citenamefont {Rigol}, \citenamefont {Bryant},\
  and\ \citenamefont {Singh}}]{NLC2}%
  \BibitemOpen
  \bibfield  {author} {\bibinfo {author} {\bibfnamefont {M.}~\bibnamefont
  {Rigol}}, \bibinfo {author} {\bibfnamefont {T.}~\bibnamefont {Bryant}}, \
  and\ \bibinfo {author} {\bibfnamefont {R.~R.~P.}\ \bibnamefont {Singh}},\
  }\href@noop {} {\bibfield  {journal} {\bibinfo  {journal} {Phys. Rev. E}\
  }\textbf {\bibinfo {volume} {75}},\ \bibinfo {pages} {061118} (\bibinfo
  {year} {2007}{\natexlab{a}})}\BibitemShut {NoStop}%
\bibitem [{\citenamefont {Rigol}\ \emph
  {et~al.}(2007{\natexlab{b}})\citenamefont {Rigol}, \citenamefont {Bryant},\
  and\ \citenamefont {Singh}}]{NLC3}%
  \BibitemOpen
  \bibfield  {author} {\bibinfo {author} {\bibfnamefont {M.}~\bibnamefont
  {Rigol}}, \bibinfo {author} {\bibfnamefont {T.}~\bibnamefont {Bryant}}, \
  and\ \bibinfo {author} {\bibfnamefont {R.~R.~P.}\ \bibnamefont {Singh}},\
  }\href@noop {} {\bibfield  {journal} {\bibinfo  {journal} {Phys. Rev. E}\
  }\textbf {\bibinfo {volume} {75}},\ \bibinfo {pages} {061119} (\bibinfo
  {year} {2007}{\natexlab{b}})}\BibitemShut {NoStop}%
\bibitem [{\citenamefont {Tang}\ \emph {et~al.}(2012)\citenamefont {Tang},
  \citenamefont {Khatami},\ and\ \citenamefont {Rigol}}]{NLC4}%
  \BibitemOpen
  \bibfield  {author} {\bibinfo {author} {\bibfnamefont {B.}~\bibnamefont
  {Tang}}, \bibinfo {author} {\bibfnamefont {E.}~\bibnamefont {Khatami}}, \
  and\ \bibinfo {author} {\bibfnamefont {M.}~\bibnamefont {Rigol}},\
  }\href@noop {} {\  (\bibinfo {year} {2012})},\ \Eprint
  {http://arxiv.org/abs/arXiv:1207.3366} {arXiv:1207.3366} \BibitemShut
  {NoStop}%
\bibitem [{\citenamefont {White}(1992)}]{White92}%
  \BibitemOpen
  \bibfield  {author} {\bibinfo {author} {\bibfnamefont {S.~R.}\ \bibnamefont
  {White}},\ }\href {\doibase 10.1103/PhysRevLett.69.2863} {\bibfield
  {journal} {\bibinfo  {journal} {Phys. Rev. Lett.}\ }\textbf {\bibinfo
  {volume} {69}},\ \bibinfo {pages} {2863} (\bibinfo {year}
  {1992})}\BibitemShut {NoStop}%
\bibitem [{\citenamefont {Schollw\"ock}(2005)}]{Scholl05}%
  \BibitemOpen
  \bibfield  {author} {\bibinfo {author} {\bibfnamefont {U.}~\bibnamefont
  {Schollw\"ock}},\ }\href {\doibase 10.1103/RevModPhys.77.259} {\bibfield
  {journal} {\bibinfo  {journal} {Rev. Mod. Phys.}\ }\textbf {\bibinfo {volume}
  {77}},\ \bibinfo {pages} {259} (\bibinfo {year} {2005})}\BibitemShut
  {NoStop}%
\bibitem [{\citenamefont {Stoudenmire}\ and\ \citenamefont
  {White}(2012)}]{Stoudenmire}%
  \BibitemOpen
  \bibfield  {author} {\bibinfo {author} {\bibfnamefont {E.~M.}\ \bibnamefont
  {Stoudenmire}}\ and\ \bibinfo {author} {\bibfnamefont {S.~R.}\ \bibnamefont
  {White}},\ }\href@noop {} {\bibfield  {journal} {\bibinfo  {journal} {Annual
  Review of Condensed Matter Physics}\ }\textbf {\bibinfo {volume} {3}},\
  \bibinfo {pages} {111} (\bibinfo {year} {2012})}\BibitemShut {NoStop}%
\bibitem [{\citenamefont {Renyi}(1961)}]{A_renyi}%
  \BibitemOpen
  \bibfield  {author} {\bibinfo {author} {\bibfnamefont {A.}~\bibnamefont
  {Renyi}},\ }\href@noop {} {\bibfield  {journal} {\bibinfo  {journal} {Proc.
  of the 4th Berkeley Symposium on Mathematics, Statistics and Probability}\
  }\textbf {\bibinfo {volume} {1960}},\ \bibinfo {pages} {547} (\bibinfo {year}
  {1961})}\BibitemShut {NoStop}%
\bibitem [{\citenamefont {Metlitski}\ and\ \citenamefont
  {Grover}(2011)}]{Max_Tarun}%
  \BibitemOpen
  \bibfield  {author} {\bibinfo {author} {\bibfnamefont {M.~A.}\ \bibnamefont
  {Metlitski}}\ and\ \bibinfo {author} {\bibfnamefont {T.}~\bibnamefont
  {Grover}},\ }\href@noop {} {\  (\bibinfo {year} {2011})},\ \Eprint
  {http://arxiv.org/abs/arXiv:1112.5166} {arXiv:1112.5166} \BibitemShut
  {NoStop}%
\bibitem [{\citenamefont {Weihong}(1997)}]{Zheng_bilayer}%
  \BibitemOpen
  \bibfield  {author} {\bibinfo {author} {\bibfnamefont {Z.}~\bibnamefont
  {Weihong}},\ }\href {\doibase 10.1103/PhysRevB.55.12267} {\bibfield
  {journal} {\bibinfo  {journal} {Phys. Rev. B}\ }\textbf {\bibinfo {volume}
  {55}},\ \bibinfo {pages} {12267} (\bibinfo {year} {1997})}\BibitemShut
  {NoStop}%
\bibitem [{\citenamefont {Hamer}\ \emph {et~al.}(2012)\citenamefont {Hamer},
  \citenamefont {Oitmaa},\ and\ \citenamefont {Weihong}}]{Hamer_bilayer}%
  \BibitemOpen
  \bibfield  {author} {\bibinfo {author} {\bibfnamefont {C.~J.}\ \bibnamefont
  {Hamer}}, \bibinfo {author} {\bibfnamefont {J.}~\bibnamefont {Oitmaa}}, \
  and\ \bibinfo {author} {\bibfnamefont {Z.}~\bibnamefont {Weihong}},\ }\href
  {\doibase 10.1103/PhysRevB.85.014432} {\bibfield  {journal} {\bibinfo
  {journal} {Phys. Rev. B}\ }\textbf {\bibinfo {volume} {85}},\ \bibinfo
  {pages} {014432} (\bibinfo {year} {2012})}\BibitemShut {NoStop}%
\bibitem [{\citenamefont {Campostrini}\ \emph {et~al.}(2002)\citenamefont
  {Campostrini}, \citenamefont {Hasenbusch}, \citenamefont {Pelissetto},
  \citenamefont {Rossi},\ and\ \citenamefont {Vicari}}]{Campo}%
  \BibitemOpen
  \bibfield  {author} {\bibinfo {author} {\bibfnamefont {M.}~\bibnamefont
  {Campostrini}}, \bibinfo {author} {\bibfnamefont {M.}~\bibnamefont
  {Hasenbusch}}, \bibinfo {author} {\bibfnamefont {A.}~\bibnamefont
  {Pelissetto}}, \bibinfo {author} {\bibfnamefont {P.}~\bibnamefont {Rossi}}, \
  and\ \bibinfo {author} {\bibfnamefont {E.}~\bibnamefont {Vicari}},\ }\href
  {\doibase 10.1103/PhysRevB.65.144520} {\bibfield  {journal} {\bibinfo
  {journal} {Phys. Rev. B}\ }\textbf {\bibinfo {volume} {65}},\ \bibinfo
  {pages} {144520} (\bibinfo {year} {2002})}\BibitemShut {NoStop}%
\bibitem [{\citenamefont {Zinn-Justin}(2002)}]{ZJ}%
  \BibitemOpen
  \bibfield  {author} {\bibinfo {author} {\bibfnamefont {J.}~\bibnamefont
  {Zinn-Justin}},\ }\href@noop {} {\emph {\bibinfo {title} {Quantum field
  theory and critical phenomena}}}\ (\bibinfo  {publisher} {Clarendon Press,
  Oxford},\ \bibinfo {year} {2002})\BibitemShut {NoStop}%
\bibitem [{\citenamefont {Chubukov}\ and\ \citenamefont
  {Morr}(1995)}]{Chubukov_bilayer}%
  \BibitemOpen
  \bibfield  {author} {\bibinfo {author} {\bibfnamefont {A.~V.}\ \bibnamefont
  {Chubukov}}\ and\ \bibinfo {author} {\bibfnamefont {D.~K.}\ \bibnamefont
  {Morr}},\ }\href {\doibase 10.1103/PhysRevB.52.3521} {\bibfield  {journal}
  {\bibinfo  {journal} {Phys. Rev. B}\ }\textbf {\bibinfo {volume} {52}},\
  \bibinfo {pages} {3521} (\bibinfo {year} {1995})}\BibitemShut {NoStop}%
\bibitem [{\citenamefont {Shevchenko}\ and\ \citenamefont
  {Sushkov}(1999)}]{bilayer_SWT1}%
  \BibitemOpen
  \bibfield  {author} {\bibinfo {author} {\bibfnamefont {P.~V.}\ \bibnamefont
  {Shevchenko}}\ and\ \bibinfo {author} {\bibfnamefont {O.~P.}\ \bibnamefont
  {Sushkov}},\ }\href {\doibase 10.1103/PhysRevB.59.8383} {\bibfield  {journal}
  {\bibinfo  {journal} {Phys. Rev. B}\ }\textbf {\bibinfo {volume} {59}},\
  \bibinfo {pages} {8383} (\bibinfo {year} {1999})}\BibitemShut {NoStop}%
\bibitem [{\citenamefont {Kotov}\ \emph {et~al.}(1998)\citenamefont {Kotov},
  \citenamefont {Sushkov}, \citenamefont {Weihong},\ and\ \citenamefont
  {Oitmaa}}]{bilayer_SWT2}%
  \BibitemOpen
  \bibfield  {author} {\bibinfo {author} {\bibfnamefont {V.~N.}\ \bibnamefont
  {Kotov}}, \bibinfo {author} {\bibfnamefont {O.}~\bibnamefont {Sushkov}},
  \bibinfo {author} {\bibfnamefont {Z.}~\bibnamefont {Weihong}}, \ and\
  \bibinfo {author} {\bibfnamefont {J.}~\bibnamefont {Oitmaa}},\ }\href
  {\doibase 10.1103/PhysRevLett.80.5790} {\bibfield  {journal} {\bibinfo
  {journal} {Phys. Rev. Lett.}\ }\textbf {\bibinfo {volume} {80}},\ \bibinfo
  {pages} {5790} (\bibinfo {year} {1998})}\BibitemShut {NoStop}%
\bibitem [{\citenamefont {Sommer}\ \emph {et~al.}(2001)\citenamefont {Sommer},
  \citenamefont {Vojta},\ and\ \citenamefont {Becker}}]{Sommer}%
  \BibitemOpen
  \bibfield  {author} {\bibinfo {author} {\bibfnamefont {T.}~\bibnamefont
  {Sommer}}, \bibinfo {author} {\bibfnamefont {M.}~\bibnamefont {Vojta}}, \
  and\ \bibinfo {author} {\bibfnamefont {K.~W.}\ \bibnamefont {Becker}},\
  }\href@noop {} {\bibfield  {journal} {\bibinfo  {journal} {Eur. Phys. J. B}\
  }\textbf {\bibinfo {volume} {23}},\ \bibinfo {pages} {329} (\bibinfo {year}
  {2001})}\BibitemShut {NoStop}%
\bibitem [{\citenamefont {Guennebaud}\ \emph {et~al.}(2010)\citenamefont
  {Guennebaud}, \citenamefont {Jacob} \emph {et~al.}}]{eigenweb}%
  \BibitemOpen
  \bibfield  {author} {\bibinfo {author} {\bibfnamefont {G.}~\bibnamefont
  {Guennebaud}}, \bibinfo {author} {\bibfnamefont {B.}~\bibnamefont {Jacob}},
  \emph {et~al.},\ }\href@noop {} {\enquote {\bibinfo {title} {Eigen v3},}\
  }\bibinfo {howpublished} {http://eigen.tuxfamily.org} (\bibinfo {year}
  {2010})\BibitemShut {NoStop}%
\bibitem [{\citenamefont {Schollw\"ock}(2011)}]{Schollwoeck:2011}%
  \BibitemOpen
  \bibfield  {author} {\bibinfo {author} {\bibfnamefont {U.}~\bibnamefont
  {Schollw\"ock}},\ }\href@noop {} {\bibfield  {journal} {\bibinfo  {journal}
  {Annals of Physics}\ }\textbf {\bibinfo {volume} {326}},\ \bibinfo {pages}
  {96} (\bibinfo {year} {2011})}\BibitemShut {NoStop}%
\bibitem [{\citenamefont {Sen}\ and\ \citenamefont {Sandvik}(2014)}]{AndersPC}%
  \BibitemOpen
  \bibfield  {author} {\bibinfo {author} {\bibfnamefont {A.}~\bibnamefont
  {Sen}}\ and\ \bibinfo {author} {\bibfnamefont {A.~W.}\ \bibnamefont
  {Sandvik}},\ }\href@noop {} {\bibfield  {journal} {\bibinfo  {journal}
  {unpublished}\ } (\bibinfo {year} {2014})}\BibitemShut {NoStop}%
\bibitem [{\citenamefont {Humeniuk}\ and\ \citenamefont
  {Roscilde}(2012)}]{Tommaso}%
  \BibitemOpen
  \bibfield  {author} {\bibinfo {author} {\bibfnamefont {S.}~\bibnamefont
  {Humeniuk}}\ and\ \bibinfo {author} {\bibfnamefont {T.}~\bibnamefont
  {Roscilde}},\ }\href@noop {} {\bibfield  {journal} {\bibinfo  {journal}
  {Phys. Rev. B}\ }\textbf {\bibinfo {volume} {86}},\ \bibinfo {pages} {235116}
  (\bibinfo {year} {2012})}\BibitemShut {NoStop}%
\bibitem [{\citenamefont {Turner}\ \emph {et~al.}(2011)\citenamefont {Turner},
  \citenamefont {Pollmann},\ and\ \citenamefont {Berg}}]{Ari}%
  \BibitemOpen
  \bibfield  {author} {\bibinfo {author} {\bibfnamefont {A.~M.}\ \bibnamefont
  {Turner}}, \bibinfo {author} {\bibfnamefont {F.}~\bibnamefont {Pollmann}}, \
  and\ \bibinfo {author} {\bibfnamefont {E.}~\bibnamefont {Berg}},\ }\href
  {\doibase 10.1103/PhysRevB.83.075102} {\bibfield  {journal} {\bibinfo
  {journal} {Phys. Rev. B}\ }\textbf {\bibinfo {volume} {83}},\ \bibinfo
  {pages} {075102} (\bibinfo {year} {2011})}\BibitemShut {NoStop}%
\bibitem [{\citenamefont {Brown}\ and\ \citenamefont {Henneaux}(1986)}]{Brown}%
  \BibitemOpen
  \bibfield  {author} {\bibinfo {author} {\bibfnamefont {J.~D.}\ \bibnamefont
  {Brown}}\ and\ \bibinfo {author} {\bibfnamefont {M.}~\bibnamefont
  {Henneaux}},\ }\href {\doibase 10.1007/BF01211590} {\bibfield  {journal}
  {\bibinfo  {journal} {Commun.Math.Phys.}\ }\textbf {\bibinfo {volume}
  {104}},\ \bibinfo {pages} {207} (\bibinfo {year} {1986})}\BibitemShut
  {NoStop}%
\bibitem [{\citenamefont {Sandvik}(2007)}]{JQ}%
  \BibitemOpen
  \bibfield  {author} {\bibinfo {author} {\bibfnamefont {A.~W.}\ \bibnamefont
  {Sandvik}},\ }\href {\doibase 10.1103/PhysRevLett.98.227202} {\bibfield
  {journal} {\bibinfo  {journal} {Phys. Rev. Lett.}\ }\textbf {\bibinfo
  {volume} {98}},\ \bibinfo {pages} {227202} (\bibinfo {year}
  {2007})}\BibitemShut {NoStop}%
\bibitem [{\citenamefont {Senthil}\ \emph {et~al.}(2004)\citenamefont
  {Senthil}, \citenamefont {Vishwanath}, \citenamefont {Balents}, \citenamefont
  {Sachdev},\ and\ \citenamefont {Fisher}}]{DQCP}%
  \BibitemOpen
  \bibfield  {author} {\bibinfo {author} {\bibfnamefont {T.}~\bibnamefont
  {Senthil}}, \bibinfo {author} {\bibfnamefont {A.}~\bibnamefont {Vishwanath}},
  \bibinfo {author} {\bibfnamefont {L.}~\bibnamefont {Balents}}, \bibinfo
  {author} {\bibfnamefont {S.}~\bibnamefont {Sachdev}}, \ and\ \bibinfo
  {author} {\bibfnamefont {M.~P.~A.}\ \bibnamefont {Fisher}},\ }\href {\doibase
  10.1126/science.1091806} {\bibfield  {journal} {\bibinfo  {journal}
  {Science}\ }\textbf {\bibinfo {volume} {303}},\ \bibinfo {pages} {1490}
  (\bibinfo {year} {2004})}\BibitemShut {NoStop}%
\bibitem [{\citenamefont {Kaul}\ \emph {et~al.}(2013)\citenamefont {Kaul},
  \citenamefont {Melko},\ and\ \citenamefont {Sandvik}}]{ARCMP_Kaul}%
  \BibitemOpen
  \bibfield  {author} {\bibinfo {author} {\bibfnamefont {R.~K.}\ \bibnamefont
  {Kaul}}, \bibinfo {author} {\bibfnamefont {R.~G.}\ \bibnamefont {Melko}}, \
  and\ \bibinfo {author} {\bibfnamefont {A.~W.}\ \bibnamefont {Sandvik}},\
  }\href@noop {} {\bibfield  {journal} {\bibinfo  {journal} {Annu. Rev. Con.
  Mat. Phys.}\ }\textbf {\bibinfo {volume} {4}},\ \bibinfo {pages} {179}
  (\bibinfo {year} {2013})}\BibitemShut {NoStop}%
\end{thebibliography}%

\end{document}